\documentclass[aps,prc,amsfonts,superscriptaddress,showpacs,
twocolumn,byrevtex
,nofootinbib
]{revtex4}

\usepackage{enumerate}
\usepackage{epsfig}
\usepackage{amsmath}
\usepackage{amssymb}
\usepackage{amsfonts}
\usepackage{color}

\allowdisplaybreaks     

\newcommand{\JMcomm}[1]{{\textcolor{black}{ #1}}}

\newcommand{\old}[1]{{\textcolor{black}{ }}}

\renewcommand{\vec}[1]{\mathbf{#1}}

\begin{document}

\title{Generalization of internal Density Functional Theory and Kohn-Sham scheme to multicomponent \JMcomm{self-bound} systems
and link with traditional DFT.}

\author{J{\'e}r{\'e}mie Messud}
%
\affiliation{Universit{\'e} Bordeaux, CNRS/IN2P3,
             Centre d'Etudes Nucl{\'e}aires de Bordeaux Gradignan, UMR5797,
             F-33175 Gradignan, France}
\affiliation{Conseil R\'egional d'Aquitaine, F-33077 Bordeaux, France}
%
%
\date{\today} 
\begin{abstract}
We generalize the recently developped "internal" Density Functional Theory (DFT) and Kohn-Sham scheme
to multicomponent systems.
We obtain a general formalism, applicable for the description of multicomponent self-bound systems
(such as \JMcomm{molecular systems} where the nuclei are treated explicitly, atomic nuclei and mixtures of $^3$He and $^4$He droplets),
where the fundamental translational symmetry has been treated correctly.
The main difference with traditional DFT is the explicit inclusion 
of center-of-mass correlations in the functional.
A large part of the paper is dedicated to the application to molecular systems,
which permits us to clarify the approximations that underly traditional DFT.
\end{abstract}


\maketitle 

%
%
\section{Introduction}

Density Functional Theory (DFT) \cite{Dre90,Koh99,DFTLN} is a tool widely used 
in condensed-matter physics and quantum chemistry to calculate 
properties of many-electrons systems.
It is based on the simple local density
instead of the less tractable $N$-body wave function. 
One of the pillars of DFT is the Hohenberg-Kohn (HK) theorem \cite{Hoh64}, 
which, in its original form, proves that for any non-degenerate system 
of Fermions or Bosons \cite{Dre90} put into a local external potential $v_{ext}(\mathbf{r})$, there exists 
a unique functional of the local one-body density $\rho(\vec{r})$ that 
gives the exact ground-state energy when $\rho(\vec{r})$ corresponds to 
the exact ground-state density.
The Kohn-Sham (KS) scheme provides a practical
method to compute self-consistently the ground-state density in a quantum 
framework, defining the local single-particle potential (i.e.\ an auxiliary 
non-interacting system) which reproduces the exact ground-state density \cite{Koh65}.

Traditional DFT is particularly well suited to describe the electrons in a molecule,
but not for the description of self-bound systems, such as atomic nuclei, He droplets or molecular systems where the nuclei are treated explicitly.
Indeed, in those systems, external fields are not required to obtain bound states.
As a consequence, isolated self-bound systems are plagued by a center-of-mass (c.m.) problem:
for any stationary state, the c.m.\ is delocalized in the whole space (because of the translational invariance of the Hamiltonian)
and the laboratory wave function is consequently not normalizable.
This prevents us from using traditional DFT 
\JMcomm{methods, formulated in terms of the laboratory density,} 
when $v_{ext}(\mathbf{r}) = 0$~\cite{Mes09}.
\JMcomm{Indeed, the laboratory density is then an indeterminate constant \cite{Eng07,Kre01,Mes09}, which prevents us to construct from it a universal
functional.}
Moreover, it is internal properties (measured in the c.m. frame) and not laboratory properties that are of experimental interest.
Experimentalists always deduce internal properties 
using the c.m. observables (position, momentum or kinetic energy of the c.m.)
and  Galilean invariance to transform all the other observables into the c.m. frame.

It is thus a question of interest to formulate a rigorous DFT formalism and KS scheme
in terms of the internal density $\rho_{int}$,
%
%
having achieved the correct separation of internal properties from the c.m.\ motion.
Efforts in that direction have been made recently in Refs.~\cite{Eng07,Bar07},
\JMcomm{
but the question of a rigorous "internal" KS scheme remained open
(see the corresponding discussion of Ref.~\cite{Mes09} for more details).
The different approach found in Ref.~\cite{Gir08b,Gir08c}
results in a rigorous KS scheme, but that is not "internal" (i.e. formulated in the c.m. frame of reference), 
so that it is not directly comparable to self-consistent 
mean-field like calculations with effective interactions formulated in the c.m. frame of reference.
}
In Ref.~\cite{Mes09}, it was proposed to use Jacobi coordinates
to decouple the c.m. properties from internal ones in self-bound systems,
which permits one to separate the non-normalizable part of the wave function from the normalizable part which describes internal properties.
Moreover, an arbitrary translationally invariant potential of the form $\sum_{i=1}^N v_{int}(\mathbf{r}_i - \mathbf{R})$, where $\vec{R}=\frac{1}{N}\sum_{j=1}^{N}\vec{r}_j$ is the total c.m.\ of the particles, was added to the Hamiltonian of a self-bound system.
This potential is an "internal" potential, i.e. it acts in the c.m. frame,
and is the only form that satisfies all the key formal properties \cite{Mes09}.
Of course $v_{int}$ should be zero in the stationary isolated self-bound case.
This is why in Ref.~\cite{Mes09} the authors presented it as a mathematical "auxiliary"
to reach the desired goal and showed that it can be dropped properly at the end, preserving all the conclusions
\JMcomm{(because the internal ground state should by definition remain bound at this limit)}.
Nevertheless, its form is suitable to model internal effects of 
fields used in experiments.
Through it and using Jacobi coordinates it was shown, by a different way than those found in Refs.~\cite{Eng07,Bar07},
the stationary "internal" DFT theorem for identical particles:
the internal many-body \JMcomm{ground} state can be written as a functional of the internal density $\rho_{int}$.
Then, the corresponding "internal" KS scheme was formulated rigorously \textit{in the c.m.\ frame}.
This work provided a first step towards a fundamental justification for the use of internal density functionals
for stationary mean-field like calculations of nuclei \cite{Ben03} or He droplets \cite{Bar06} with effective interactions,
showing that there exists an ultimate functional which permits one to reproduce the exact internal density.
The major difference with traditional DFT is that the c.m. correlations (which appear due to the redundant coordinate problem) 
are explicitly included in the functional.

The aim of the present article is to generalize the internal DFT formalism and KS scheme to self-bound systems composed of different kinds of particles
\JMcomm{(i.e. multicomponent systems)}.
This is crucial to obtain a fundamental justification to the use of density functionals for the description of:
\begin{enumerate}[(i)]
\item \textit{Molecular systems where the nuclei are treated explicitly and quantum mechanically.}\\
Taking into account the quantum nature of the nuclei is important to describe small molecules \cite{Gro95} and solid hydrogen \cite{Sur93} for instance,
and explicit treatment of the nuclei is necessary to decribe non-adiabatic phenomena.
Moreover, taking into account the electrons-nuclei correlations can be important for the description of some physical properties \cite{Cha08}.
For those reasons, various multicomponent DFT formalisms have been developped \cite{Cap82,Gid98,Kre01,Kre08,Cha08,Cha09}.
But in none of them has the c.m. motion been separated properly \JMcomm{from the beginning}, 
leading to formal difficulties (that will be detailed in \S\ref{par:link1}).
Here, we propose a general formulation that overcomes those difficulties and is suited for the description of all self-bound systems.
We show that the application to molecular systems is particularly interesting for understanding the approximations inherent to traditional DFT
and thus how to eventually improve its results for molecular systems.
\item \textit{Atomic nuclei.}\\
The application of multicomponent internal DFT to protons and neutrons gives a further step towards the use of 
mean-field like calculations with effective interactions, as currently done in nuclear physics \cite{Ben03}, and shows that the c.m. 
correlations can be included in the functional. It thus opens the way to an alternative to the numerically very costly projection techniques
used in nuclear physics to restore Galilean invariance \cite{Schm01a,Schm04,Pei62,Ben03}:
they wouldn't be necessary if the ultimate functional were known.
\item \textit{Mixtures of $^3$He and $^4$He droplets.}\\
Mean-field like calculations with effective interactions are also used to describe those systems \cite{Bar06}
and can be justified more fundamentally by the multicomponent internal DFT formalism.
\end{enumerate}

The article is organized as follows. We first give the foundations of the multicomponent internal DFT formalism and
make explicit the link with the previously developped formalisms.
Then, we apply the formalism to molecular systems where the nuclei are treated explicitly.
This application will show clearly the link between multicomponent internal DFT 
and "one kind of particle" standard and internal DFT.
Finally, we mention some features of the application 
to atomic nuclei and mixtures of $^3$He and $^4$He droplets.

%
%

\section{Many body formulation.}

\subsection{General formulation.}
\label{par:gen_form}

For the sake of simplicity and to underline more clearly the physics, we assume:
\begin{enumerate}[(i)]
\item \textit{Two different kinds of particles}.\\
This is sufficient to describe atomic nuclei, mixtures of $^3$He and $^4$He droplets
and molecular systems with only one kind of nuclei.
The generalization to more kinds of different particles, for the description of molecular systems
with more than one kind of nuclei, is easy.
\item \textit{Two-body particle-particle interactions}.\\The generalization to 3-body etc interactions is straightforward.
\item \textit{Particles without spin}.\\
Generalization to spin polarized systems can be obtained by introduction of an arbitrary \textit{internal} 
magnetic field $\sum_{i} \mathbf{B}_{int}(\mathbf{r}_i - \mathbf{R})$,
which models the effect on internal properties of magnetic fields used in experiments,
and adapting the standard derivation \cite{Dre90,Bar72}.
\end{enumerate}
We thus consider:
\begin{enumerate}[(i)]
\item $N^{(1)}$ particles $(1)$ of mass $m^{(1)}$ and laboratory coordinates $\{\vec{r}^{(1)}_1,\dots,\vec{r}^{(1)}_{N^{(1)}}\}$,
\item $N^{(2)}$ particles $(2)$ of mass $m^{(2)}$ and laboratory coordinates $\{\vec{r}^{(2)}_1,\dots,\vec{r}^{(2)}_{N^{(2)}}\}$,
\end{enumerate}
and start from a general translationally invariant many-body Hamiltonian
\footnote{
In a Hamiltonian and wave function based 
description of an isolated self-bound system, the Hamiltonian should 
be explicitly translationally invariant to ensure Galilean invariance 
of the wave function.
Translational invariance, which states that the observables do not depend 
on the position of the c.m., is a necessary but not sufficient condition 
for the more fundamental Galilean invariance, which ensures that scalar observables
are the same in all inertial frames.
}

\begin{eqnarray}
\label{eq:H}
H
&=&   \sum_{i=1}^{N^{(1)}} \frac{{\vec{p}_i^{(1)}}^2}{2m^{(1)}} + \sum_{i=1}^{N^{(2)}} \frac{{\vec{p}_i^{(2)}}^2}{2m^{(2)}}
\nonumber\\
  &&+ \sum_{\stackrel{i,j=1}{i > j}}^{N^{(1)}} u^{(1)} (\vec{r}^{(1)}_i-\vec{r}^{(1)}_j) 
  + \sum_{\stackrel{i,j=1}{i > j}}^{N^{(2)}} u^{(2)} (\vec{r}^{(2)}_i-\vec{r}^{(2)}_j) 
\nonumber\\
  &&  + \sum_{i=1}^{N^{(1)}}\sum_{j=1}^{N^{(2)}} u^{(12)} (\vec{r}^{(1)}_i-\vec{r}^{(2)}_j) 
\nonumber\\
  &&+ \sum_{i=1}^{N^{(1)}} v^{(1)}_{\text{int}} (\vec{r}^{(1)}_i - \vec{R})
  + \sum_{i=1}^{N^{(2)}} v^{(2)}_{\text{int}} (\vec{r}^{(2)}_i - \vec{R})
,
\end{eqnarray}
composed of:
\begin{enumerate}[(i)]
\item the usual kinetic energy terms, 
\item translationally invariant 
two-body potentials $u^{(1)}$,  $u^{(2)}$ and $u^{(12)}$ which describe the particle-particle 
interactions,
\item arbitrary translationally invariant "internal" potentials $v_{int}^{(1)}$ and $v_{int}^{(2)}$,
which act on each species in the c.m. frame and can model internal effects of potentials used in 
experiments (such as electric fields in the molecular case), or
can be safely dropped at the end in the case of an isolated self-bound system.
\end{enumerate}

The total c.m.\ coordinate $\vec{R}$, i.e. of particles $(1)$ and $(2)$, is defined as
\begin{eqnarray}
\vec{R}&=&\frac{1}{N^{(1)} m^{(1)}+N^{(2)} m^{(2)}}
\Big[
m^{(1)} \sum_{i=1}^{N^{(1)}}\vec{r}^{(1)}_i + m^{(2)} \sum_{i=1}^{N^{(2)}}\vec{r}^{(2)}_i
\Big]
\nonumber\\
&=& \frac{M^{(1)}\mathbf{R}^{(1)} + M^{(2)}\mathbf{R}^{(2)}}{M^{(1)} + M^{(2)}}
,
\label{eq:cm}
\end{eqnarray}
where $M^{(l)}=N^{(l)} m^{(l)}$ is the total mass of the particles $(l)$ and $\mathbf{R}^{(l)}=\frac{1}{N^{(l)}}\sum_{i=1}^{N^{(l)}} \vec{r}^{(l)}_i$
is the center of mass of the particles $(l)$.
The ($N^{(1)}+N^{(2)}-1$) Jacobi coordinates $\{\mathbf{\xi}_{\alpha}\}$ are defined as
(see appendix \ref{app:jacobi2} for some reminders)
\begin{eqnarray}
\label{eq:jacobi}
&& for\hspace{3mm}\alpha\in[1;N^{(1)}-1]: \quad \mathbf{\xi}_{\alpha} = \vec{r}^{(1)}_{\alpha+1}-\frac{1}{\alpha}\sum_{i=1}^{\alpha}\vec{r}^{(1)}_i,
\\
&&for\hspace{3mm}\alpha=N^{(1)}: \quad\quad\quad\quad \mathbf{\xi}_{N^{(1)}} = \vec{r}^{(2)}_{1}-\vec{R}^{(1)},
\nonumber\\
&&for\hspace{3mm}\alpha\in[N^{(1)}+1;N^{(1)}+N^{(2)}-1]: 
\nonumber\\
&&\quad \mathbf{\xi}_{\alpha} = \vec{r}^{(2)}_{\alpha-N^{(1)}+1}-
\frac{m^{(2)}\sum_{i=1}^{\alpha-N^{(1)}}\vec{r}^{(2)}_i + 
M^{(1)}\vec{R}^{(1)}
}{(\alpha-N^{(1)}) m^{(2)}+M^{(1)}}
\nonumber.
\end{eqnarray}
The $\xi_\alpha$ are relative to the c.m.\ of the other 
$1, \ldots, \alpha-1$ particles and are independent of $\vec{R}$. They are to be distinguished from the ($N^{(1)}+N^{(2)}$)
"laboratory coordinates" $\vec{r}_i^{(l)}$, and the ($N^{(1)}+N^{(2)}$) "c.m. frame coordinates" $(\vec{r}_i^{(l)}-\vec{R})$ relative to the total c.m.\ .

It is to be noted that, for multicomponent systems, those "standard" Jacobi coordinates cannot be associated to one specific kind of particle,
in contrary to the laboratory or c.m. frame coordinates. 
More precisely, even if the $\{\mathbf{\xi}_{\alpha} ; \alpha\in[1;N^{(1)}-1] \}$ are the Jacobi coordinates associated to particles $(1)$ only, the $\{\mathbf{\xi}_{\alpha} ; \alpha\in[N^{(1)};N^{(1)}+N^{(2)}-1]\}$ mix the laboratory coordinates of particles $(1)$ and $(2)$.

We mention that the "standard" Jacobi coordinates used in this paper do not constitute the unique set of coordinates which permit to separate the c.m. contribution from the internal contribution in the Hamiltonian, as we will do thereafter.
Another possible decomposition is
(instead of (\ref{eq:jacobi}))
%
\begin{eqnarray}
&& for\hspace{3mm}l={1,2}\hspace{3mm}and\hspace{3mm}\alpha\in[1;N^{(l)}-1]: 
\nonumber\\
&& \hspace{3cm}
\mathbf{\xi}_{\alpha}^{(l)} = \vec{r}^{(l)}_{\alpha+1}-\frac{1}{\alpha}\sum_{i=1}^{\alpha}\vec{r}^{(l)}_i,
\nonumber\\
&& and\hspace{1cm} \quad \vec{R}^{(12)} = \vec{R}^{(2)} - \vec{R}^{(1)}
\label{eq:jacobi__2}
,
\end{eqnarray}
%
in addition to the total c.m. coordinate $\vec{R}$ defined in Eq.~ (\ref{eq:cm}).
$\vec{R}^{(12)}$ is the relative motion of the c.m. of each species.
The use of this set of coordinates permits one to introduce the Jacobi coordinates $\mathbf{\xi}_{\alpha}^{(l)}$ 
reduced to a given kind of particles ($l$), which is more symmetric.
In this article, we decided to use the set defined in Eq.~(\ref{eq:jacobi}) because 
it permits us to obtain more directly the results of Appendix \ref{app:jacobi3} and \S \ref{par:tradDFT1}
in the limit when one kind of particle has a much larger mass than the other kind,
and does not change the final results 
nor the way to obtain them (the formulation in terms of the set (\ref{eq:jacobi__2}) is straightforward).

Since the $\{\vec{r}^{(l)}_i-\vec{r}^{(l)}_{j\ne i}\}$, $\{\vec{r}^{(l)}_i-\vec{r}^{(m\ne l)}_{j}\}$ 
and $\{\vec{r}^{(l)}_i - \vec{R}\}$ can be rewritten as functions of the $\xi_\alpha$,
the interaction potentials $u^{(1)}$, $u^{(2)}$, $u^{(12)}$, and the internal potentials $v^{(1)}_{int}$ and $v^{(2)}_{int}$ 
can be rewritten as functions of the $\xi_\alpha$. We denote: 
\begin{eqnarray}
&&\sum_{\stackrel{i,j=1}{i > j}}^{N^{(l)}} u^{(l)} (\vec{r}^{(l)}_i-\vec{r}^{(l)}_j)
\quad\rightarrow\quad
U^{(l)}(\mathbf{\xi}_1, ..., \mathbf{\xi}_{N^{(1)}+N^{(2)}-1}) 
\nonumber\\
&&\sum_{i=1}^{N^{(1)}}\sum_{j=1}^{N^{(2)}} u^{(12)} (\vec{r}^{(1)}_i-\vec{r}^{(2)}_j)
\quad\rightarrow\quad
U^{(12)}(\mathbf{\xi}_1, ..., \mathbf{\xi}_{N^{(1)}+N^{(2)}-1}) 
\nonumber\\
&&\sum_{i=1}^{N^{(l)}} v^{(l)}_{int} (\vec{r}^{(l)}_i - \vec{R})
\quad\rightarrow\quad
V^{(l)}_{int}(\mathbf{\xi}_1, ..., \mathbf{\xi}_{N^{(1)}+N^{(2)}-1})
.
\label{eq:V_int}
\end{eqnarray}
After having defined the conjugate momenta of $\vec{R}$ and $\xi_\alpha$ (see appendix \ref{app:jacobi2}),
we can separate (\ref{eq:H}) into $H = H_\text{CM} + H_\text{int}$, where the c.m. Hamiltonian ($M = N^{(1)}m^{(1)}+N^{(2)}m^{(2)}$ is the total mass)

\begin{equation}
H_\text{CM} = -\frac{\hbar^2 \Delta_\vec{R}}{2M} 
\label{eq:H_cm}
\end{equation}
is a one-body operator acting in  $\vec{R}$ space only, and the internal Hamiltonian
\begin{eqnarray}
\label{eq:H_int}
&&H_\text{int}=\sum_{\alpha=1}^{N^{(1)}+N^{(2)}-1} \frac{\tau_\alpha^2}{2\mu_\alpha}
\nonumber\\
&&\quad+U^{(1)}(\mathbf{\xi}_1, ..., \mathbf{\xi}_{N^{(1)}+N^{(2)}-1})
+U^{(2)}(\mathbf{\xi}_1, ..., \mathbf{\xi}_{N^{(1)}+N^{(2)}-1})
\nonumber\\
&&\quad+U^{(12)}(\mathbf{\xi}_1, ..., \mathbf{\xi}_{N^{(1)}+N^{(2)}-1})
\nonumber\\
&&\quad+V^{(1)}_{int}(\mathbf{\xi}_1, ..., \mathbf{\xi}_{N^{(1)}+N^{(2)}-1})
+V^{(2)}_{int}(\mathbf{\xi}_1, ..., \mathbf{\xi}_{N^{(1)}+N^{(2)}-1})
\nonumber
\end{eqnarray}
is a $(N^{(1)}+N^{(2)}-1)$ body operator in the $\{\xi_\alpha\}$ space.
It contains the particle-particle interaction potentials and the internal potentials.
$\tau_\alpha$ is the conjugate momentum of $\xi_\alpha$ and $\mu_\alpha$ the corresponding reduced mass, defined as:
\begin{eqnarray}
&& for\hspace{3mm}\alpha\in[1;N^{(1)}-1]: \quad
\mu_\alpha = \frac{\alpha}{\alpha+1}m^{(1)}
,
\nonumber\\
&&for\hspace{3mm}\alpha\in[N^{(1)};N^{(1)}+N^{(2)}-1]:
\nonumber\\
&& \quad\quad\quad \mu_\alpha = \frac{ \big[ M^{(1)} + (\alpha - N^{(1)}) m^{(2)} \big] m^{(2)} }{M^{(1)}+(\alpha-N^{(1)}+1)m^{(2)}}
.
\label{eq:red_mass}
\end{eqnarray}

Hence, the ($N^{(1)}+N^{(2)}$)-body  laboratory wave function 
$\psi(\vec{r}^{(1)}_1, \ldots , \vec{r}^{(1)}_{N^{(1)}} ; \vec{r}^{(2)}_{1}, \ldots , \vec{r}^{(2)}_{N^{(2)}})$
can be separated into a wave function 
$\Gamma$ that is an eigenstate of $H_\text{CM}$ and depends on the c.m. coordinate $\vec{R}$ only, and an "internal" wave function 
$\psi_{int}$ that is an eigenstate of $H_\text{int}$ and depends on the remaining ($N^{(1)}+N^{(2)}-1$) Jacobi 
coordinates ${\boldmath{\xi}}_\alpha$:
\begin{eqnarray}
\label{eq:psi}
\lefteqn{\psi(\vec{r}^{(1)}_1, \ldots , \vec{r}^{(1)}_{N^{(1)}} ; \vec{r}^{(2)}_{1}, \ldots , \vec{r}^{(2)}_{N^{(2)}}) =}
\\
&&\quad\quad\quad\Gamma(\vec{R}) \; 
  \psi_{int} ({\boldmath{\xi}}_1, \ldots , {\boldmath{\xi}}_{N^{(1)}+N^{(2)}-1})
\; .
\nonumber
\end{eqnarray}
The $\Gamma (\vec{R})$ describes the motion of the c.m. of the isolated system
in any chosen inertial frame of reference (such as the laboratory).
The $\psi_{int}$ describes the internal properties and is a function of the ($N^{(1)}+N^{(2)}-1$) Jacobi 
coordinates. Of course it could also be written as a function of the $N^{(1)}$ coordinates $\mathbf{r}^{(1)}_i$ and
$N^{(2)}$ coordinates $\mathbf{r}^{(2)}_i$, 
i.e. $\psi_{int}(\vec{r}^{(1)}_1, \ldots , \vec{r}^{(1)}_{N^{(1)}} ; \vec{r}^{(2)}_{1}, \ldots , \vec{r}^{(2)}_{N^{(2)}})$,
but one of them would be redundant \cite{Die96}.

Thus, c.m.\ properties 
and internal properties
are fully decoupled and the total energy splits into $E = E_{CM}+E_{int}$.
Since $\Gamma$ is solution of the free Schr\"odinger equation, $\Gamma(\vec{R})$ should be an arbitrary 
stationary plane wave, i.e. infinitely spread and thus not normalizable.
This leads to  delocalization of $\vec{R}$ and arbitrary c.m.\ energy 
$E_{CM} = \hbar^2\mathbf{K}^2/(2M)$.
This does not correspond to experimental situations, where the system is no longer isolated: interactions with other systems
of the experimental apparatus localize the c.m..
But the formal decoupling between c.m.\ motion and internal properties 
permits us to \JMcomm{left} the c.m.\ motion to the choice of experimental conditions,
internal properties
being comparable to the experimental ones.
\JMcomm{
If $\psi_{int}$ is normalizable, which is by definition always
the case for the ground state of a self-bound system,}
the internal energy can be written
%
\begin{eqnarray}
\label{eq:E0}
E_{int}[\psi_{int}] = (\psi_{int}| H_{int} |\psi_{int})
.
\end{eqnarray}

\subsection{Some useful definitions.}
\label{par:def}

Before we adress the generalization of internal DFT to many kinds of particles,
we define some quantities and relations that will be useful for the subsequent considerations.
Following Refs.~\cite{Mes09,Gir08b,Kaz86}, we define the internal one-body densities
for each species
\begin{eqnarray}
\label{eq:rho_int0}
\rho^{(l)}_{int}(\vec{r})
   &=&  N^{(l)} \int \! d\vec{r}^{(1)}_1 \cdots d\vec{r}^{(1)}_{N^{(1)}} d\vec{r}^{(2)}_1 \cdots d\vec{r}^{(2)}_{N^{(2)}} \;
\\
    &&\times\delta(\mathbf{R})
      |\psi_{int}(\vec{r}^{(1)}_1, \ldots , \vec{r}^{(1)}_{N^{(1)}} ; \vec{r}^{(2)}_{1}, \ldots , \vec{r}^{(2)}_{N^{(2)}})|^2
\nonumber\\
    &&\times\delta \big( \vec{r} - (\vec{r}^{(l)}_i-\mathbf{R}) \big)\,
\nonumber
.
\end{eqnarray}
They are normalized to $N^{(l)}$.
The laboratory densities $\rho^{(l)}(\mathbf{r})$ are obtained by convolution of $\rho^{(l)}_{int}$ with the c.m.\ wave 
function (following \cite{Gir08b,Kaz86}):
$
\rho^{(l)}(\mathbf{r}) = \int d\mathbf{R} |\Gamma(\mathbf{R})|^2 \rho^{(l)}_{int}(\mathbf{r} - \mathbf{R})
$.

Following the considerations of Refs.~\cite{Mes09,Kaz86}, we define the local parts of the two-body internal density matrices
for each species
\begin{eqnarray}
\gamma^{(l)}_{int}(\vec{r},\vec{r'})
& = & N^{(l)}(N^{(l)}-1) \int \! d\vec{r}^{(1)}_1 \cdots d\vec{r}^{(1)}_{N^{(1)}} d\vec{r}^{(2)}_1 \cdots d\vec{r}^{(2)}_{N^{(2)}}\;
\nonumber\\
      &&\times\delta(\mathbf{R}) |\psi_{int}(\vec{r}^{(1)}_1, \ldots , \vec{r}^{(1)}_{N^{(1)}} ; \vec{r}^{(2)}_{1}, \ldots , \vec{r}^{(2)}_{N^{(2)}})|^2 \times \,
\nonumber\\
      &&\times \big( \vec{r} - (\vec{r}^{(l)}_i-\mathbf{R}) \big)
      \delta \big( \vec{r'} - (\vec{r}^{(l)}_{j\ne i}-\mathbf{R}) \big)
\label{eq:gamint0}
.
\end{eqnarray}
These have the required normalisation to $N^{(l)}(N^{(l)}-1)$.
Following similar steps to those in Refs.~\cite{Gir08b,Kaz86},
we can show that the local part of the two-body laboratory density matrices 
$\gamma^{(l)}(\vec{r},\vec{r'})$ are obtained by convolution of $\gamma^{(l)}_{int}$ with the c.m.\ wave 
function:
$
\gamma^{(l)}(\vec{r},\vec{r'}) = \int d\mathbf{R} |\Gamma(\mathbf{R})|^2 \gamma^{(l)}_{int}(\vec{r} - \mathbf{R},\vec{r'} - \mathbf{R}) .
$

Finally, we introduce the local part of the two-body internal "coupling" density matrix
\begin{eqnarray}
\label{eq:gamint1}
\lefteqn{\gamma^{(12)}_{int}(\vec{r},\vec{r'})}
      \\
& = & N^{(1)}N^{(2)} \int \! d\vec{r}^{(1)}_1 \cdots d\vec{r}^{(1)}_{N^{(1)}} d\vec{r}^{(2)}_1 \cdots d\vec{r}^{(2)}_{N^{(2)}} \;
\nonumber\\
      &&\times\delta(\mathbf{R}) |\psi_{int}(\vec{r}^{(1)}_1, \ldots , \vec{r}^{(1)}_{N^{(1)}} ; \vec{r}^{(2)}_{1}, \ldots , \vec{r}^{(2)}_{N^{(2)}})|^2 \,
\nonumber\\
      &&\times\delta \big( \vec{r} - (\vec{r}^{(1)}_i-\mathbf{R}) \big)
      \delta \big( \vec{r'} - (\vec{r}^{(2)}_{j}-\mathbf{R}) \big)
\nonumber
,
\end{eqnarray}
where $\vec{r}$ acts in the particle (1) space and $\vec{r'}$ acts in the particle (2) space.

The definitions of $\rho^{(l)}_{int}(\vec{r})$, $\gamma^{(l)}_{int}(\vec{r},\vec{r'})$ and $\gamma^{(12)}_{int}(\vec{r},\vec{r'})$ show clearly that they are defined in the c.m. frame,
i.e.\ that the positions \JMcomm{$\vec{r}$ and $\vec{r'}$} are measured in the c.m. frame (see the $\delta$ relations in Eqs. (\ref{eq:rho_int0}), (\ref{eq:gamint0}) and (\ref{eq:gamint1})).
Compared to the traditional definitions, a $\delta(\mathbf{R})$ appears in the definition of the internal densities calculated
with $\psi_{int}$ \JMcomm{written in terms of the} $\{\mathbf{r}^{(l)}_i\}$ coordinates.
Since one of them is redundant, the $\delta(\mathbf{R})$ represents the dependence of the redundant coordinate on the others~
\footnote{
More generally, we should introduce a $\delta(\mathbf{R}-\mathbf{a})$, where
$\mathbf{a}$ is an arbitrary translation vector, instead of the $\delta(\mathbf{R})$.
This is linked to the translational invariance.
In this paper we chose $\mathbf{a}=\mathbf{0}$, which leads to perfectly equivalent results
and permits to lighten the notations.
}.
Note that, following steps similar to those in Ref.~\cite{Mes09}, one can rewrite
$\rho^{(l)}_{int}$, $\gamma^{(l)}_{int}$ and $\gamma^{(12)}_{int}$ as functions of Jacobi coordinates $\{\xi_\alpha\}$.
\newline

We establish a useful relation.
For any function $f(\vec{r}^{(1)}_1, \ldots , \vec{r}^{(1)}_{N^{(1)}} ; \vec{r}^{(2)}_{1}, \ldots , \vec{r}^{(2)}_{N^{(2)}})$ of laboratory coordinates
expressible in terms of Jacobi coordinates [we denote $F(\xi_1,...,\xi_{N^{(1)}+N^{(2)}-1})$ \JMcomm{and $\hat{F}$ the associated operator}] we have
\begin{widetext}
\begin{eqnarray}
\label{eq:rel}
\lefteqn{(\psi_{int}| \JMcomm{\hat{F}} |\psi_{int})}
\\
& = & 
\int \! d\mathbf{\xi}_1 \cdots  d\mathbf{\xi}_{N^{(1)}+N^{(2)}-1} F(\xi_1,...,\xi_{N^{(1)}+N^{(2)}-1}) 
\big| \psi_{int} (\xi_1,...,\xi_{N^{(1)}+N^{(2)}-1}) \big|^2
\nonumber
\\
& = & \int \! d\mathbf{R} d\mathbf{\xi}_1 \cdots  d\mathbf{\xi}_{N^{(1)}+N^{(2)}-1} \delta(\mathbf{R}) F(\xi_1,...,\xi_{N^{(1)}+N^{(2)}-1})
\big| \psi_{int} (\xi_1,...,\xi_{N^{(1)}+N^{(2)}-1}) \big|^2
\nonumber\\
& = & \int \! d\vec{r}^{(1)}_{N^{(1)}} \cdots d\vec{r}^{(1)}_{N^{(1)}} d\vec{r}^{(2)}_1 \cdots d\vec{r}^{(2)}_{N^{(2)}}
\delta(\mathbf{R}) f(\vec{r}^{(1)}_1, \ldots , \vec{r}^{(1)}_{N^{(1)}} ; \vec{r}^{(2)}_{1}, \ldots , \vec{r}^{(2)}_{N^{(2)}})
\big| \psi_{int}(\vec{r}^{(1)}_1, \ldots , \vec{r}^{(1)}_{N^{(1)}} ; \vec{r}^{(2)}_{1}, \ldots , \vec{r}^{(2)}_{N^{(2)}}) \big|^2
\nonumber
\, .
\end{eqnarray}
\end{widetext}
We see that the "internal mean values" calculated with $\psi_{int}$
expressed as a function of the ($N^{(1)}+N^{(2)}-1$) coordinates $\xi_\alpha$, can also be calculated with $\psi_{int}$
expressed as a function of the ($N^{(1)}+N^{(2)}$) coordinates 
$\vec{r}^{(l)}_i$.
Then, as above, a $\delta(\mathbf{R})$ that
represents the dependence of the redundant coordinate on the others appears.

\section{Generalization of internal DFT and Kohn-Sham scheme to many kinds of particles.}

\subsection{Hohenberg-Kohn theorem.}

We show that $\psi_{int}$ can be written as a functional of the internal densities $\rho^{(1)}_{int}$ and $\rho^{(2)}_{int}$.
The relation (\ref{eq:rel}) leads to \JMcomm{($\hat{V}^{(l)}_{int}$ is defined with (\ref{eq:V_int}))}
\begin{eqnarray}
\label{eq:Eext}
\lefteqn{ (\psi_{int}|\JMcomm{\hat{V}^{(l)}_{int}}|\psi_{int}) } 
\\
& = & \int \! d\vec{r}^{(1)}_{1} \cdots d\vec{r}^{(1)}_{N^{(1)}} d\vec{r}^{(2)}_1 \cdots d\vec{r}^{(2)}_{N^{(2)}}\delta(\mathbf{R}) 
\nonumber\\
      &&\times\sum_{i=1}^{N^{(l)}} v^{(l)}_{int}(\vec{r}^{(l)}_i - \vec{R}) 
      |\psi_{int}(\vec{r}^{(1)}_1, \ldots , \vec{r}^{(1)}_N ; \vec{r}^{(2)}_{1}, \ldots , \vec{r}^{(2)}_{N^{(2)}})|^2 \, 
      \nonumber\\
& = & \sum_{i=1}^{N^{(l)}}
      \int \! d\vec{r} \; v^{(l)}_{int}(\vec{r}) 
      \int \! d\vec{r}^{(1)}_{1} \cdots d\vec{r}^{(1)}_{N^{(1)}} d\vec{r}^{(2)}_1 \cdots d\vec{r}^{(2)}_{N^{(2)}} \; \delta(\mathbf{R})
\nonumber\\
      &&\times|\psi_{int}(\vec{r}^{(1)}_1, \ldots , \vec{r}^{(1)}_N ; \vec{r}^{(2)}_{1}, \ldots , \vec{r}^{(2)}_{N^{(2)}})|^2
      \delta \big( \vec{r}-(\vec{r}^{(l)}_i-\vec{R} ) \big) 
      \nonumber\\
& = & \sum_{i=1}^{N^{(l)}}
      \int \! d\vec{r} \; v^{(l)}_{int}(\vec{r}) \, 
      \, \frac{\rho^{(l)}_{int}(\vec{r})}{N^{(l)}} 
      \nonumber\\
& = & \int \! d\vec{r} \; v^{(l)}_{int}(\vec{r}) \, \rho^{(l)}_{int}(\vec{r})
,
\end{eqnarray}
where we used (\ref{eq:rho_int0}) to obtain the penultimate equality.
We see that the potential $v^{(l)}_{int}(\vec{r}^{(l)}_i - \vec{R})$ that is $N^{(l)}$-body 
with respect to the laboratory coordinates, becomes one body (and local) when
expressed with the c.m. frame coordinates
(remember that the $\rho^{(l)}_{int}$ are defined in the c.m. frame, cf. \S \ref{par:def}).

The internal energy 
$E_{int}[\psi_{int}]$ (\ref{eq:E0}) can be rewritten
\begin{eqnarray}
\label{eq:newE}
E_{int}[\psi_{int}] &=&
\Big(\psi_{int}\Big|\sum_{\alpha=1}^{N^{(1)}+N^{(2)}-1} \frac{\tau_\alpha^2}{2\mu_\alpha}\Big|\psi_{int}\Big)
\\
&+&\Big(\psi_{int}\Big|\JMcomm{\hat{U}^{(1)}}
+ \JMcomm{\hat{U}^{(2)}}
+\JMcomm{\hat{U}^{(12)}}
\Big|\psi_{int}\Big)
\nonumber\\
&+&\int d\vec{r} v^{(1)}_{int}(\vec{r}) \rho^{(1)}_{int}(\vec{r})
+\int d\vec{r} v^{(2)}_{int}(\vec{r}) \rho^{(2)}_{int}(\vec{r})
\nonumber
.
\end{eqnarray}
As in its definition enter
two arbitrary one-body potentials in the c.m. frame of the form 
$\int \! d\vec{r} \; v^{(l)}_{int}(\vec{r}) \, \rho^{(l)}_{int}(\vec{r})$,
and as the ground state ${\psi}_{int}$ of $H_{int}$ is obtained by minimization of ${E}_{int}$,
we can directly apply 
the usual proof of the standard HK theorem \cite{Hoh64,Dre90} generalized to many kinds of particles \cite{Kre08,Cap82}.
We thus can claim that, for a non-degenerate 
ground state ${\psi}_{int}$
and given kinds of particles $(1)$ and $(2)$ (i.e.\ given interactions $u^{(1)}$, $u^{(2)}$ and $u^{(12)}$),
\JMcomm{the ground state} ${\psi}_{int}$ can be expressed as a unique
functional of ${\rho}^{(1)}_{int}$ and ${\rho}^{(2)}_{int}$,
i.e. ${\psi}_{int}[{\rho}^{(1)}_{int},{\rho}^{(2)}_{int}]$,
and therefore also the corresponding internal energy of a self-bound 
system, i.e. ${E}_{int}[{\rho}^{(1)}_{int},{\rho}^{(2)}_{int}]$. 
%
%

As emphasized in \cite{Lie83,Mes09}, the HK theorem is valid only for arbitrary one-body potentials 
that lead to \textit{bound} ground states. As a direct consequence, 
the internal DFT formalism is valid only for potentials 
$v^{(1)}_{int}$ and $v^{(2)}_{int}$ that lead to \textit{bound} internal ground states $\psi_{int}$. 
%
%
For pure self-bound systems, described by our formalism in the limit $v^{(1)}_{int}=v^{(2)}_{int}=0$,
$\psi_{int}$ should by definition be a bound ground state,
so that the previous conclusions still hold.

%
%
\subsection{Internal Kohn-Sham scheme.}
\label{par:intKS}

Following similar steps as in~(\ref{eq:rel}), we rewrite the
interacting kinetic energy as
(we note ($\vec{r}^{(1)}_{1}, \cdots, \vec{r}^{(1)}_{N^{(1)}} ; \vec{r}^{(2)}_1, \cdots, \vec{r}^{(2)}_{N^{(2)}}$)
$\rightarrow$ ($\vec{r}^{(1)}_{1}, \cdots, \vec{r}^{(2)}_{N^{(2)}}$)
and $d\vec{r}^{(1)}_{1} \cdots d\vec{r}^{(1)}_{N^{(1)}} d\vec{r}^{(2)}_1 \cdots d\vec{r}^{(2)}_{N^{(2)}}$
$\rightarrow$ $d\vec{r}^{(1)}_{1} \cdots d\vec{r}^{(2)}_{N^{(2)}}$ for simplicity)
\begin{widetext}
\begin{eqnarray}
\label{eq:intkin}
\Big(\psi_{int}\Big|\sum_{\alpha=1}^{N^{(1)}+N^{(2)}-1} \frac{\tau_\alpha^2}{2\mu_\alpha}\Big|\psi_{int}\Big)
& = & 
\int d\mathbf{\xi}_1 ... \mathbf{\xi}_{N^{(1)}+N^{(2)}-1} \psi_{int}^*(\{\mathbf{\xi}_\alpha\})
\Big( -\frac{\hbar^2\Delta_\vec{R}}{2M} + \sum_{\alpha=1}^{N^{(1)}+N^{(2)}-1} \frac{\tau_\alpha^2}{2\mu_\alpha} \Big)
\psi_{int}(\{\mathbf{\xi}_\alpha\})
\\
& = & \int \! d\vec{r}^{(1)}_1 \cdots d\vec{r}^{(2)}_{N^{(2)}} \delta(\mathbf{R}) \,
  \psi_{int}^*(\vec{r}^{(1)}_1, \cdots, \vec{r}^{(2)}_{N^{(2)}})
  \Big( \sum_{i=1}^{N^{(1)}} \frac{\mathbf{p}^{(1)2}_i}{2m^{(1)}}
  + \sum_{i=1}^{N^{(2)}} \frac{\mathbf{p}^{(2)2}_i}{2m^{(2)}} \Big)
  \psi_{int}(\vec{r}^{(1)}_1, \cdots, \vec{r}^{(2)}_{N^{(2)}})
\nonumber
\, ,
\end{eqnarray}
\end{widetext}
which permits to recover an interpretation in terms of particles $(1)$ and $(2)$
and makes it clear that the major difference with the standard kinetic energy
comes from the c.m.\ correlations, i.e. the $\delta(\mathbf{R})$ term in the previous expression.

Using relation (\ref{eq:intkin}) and the notations introduced in \S\ref{par:def}, we can rewrite the internal energy as
\begin{widetext}
\begin{eqnarray}
\label{eq:newE2}
E_{int}[{\rho}^{(1)}_{int},{\rho}^{(2)}_{int}] &=&
\int \! d\vec{r}^{(1)}_1 \cdots d\vec{r}^{(2)}_{N^{(2)}} \delta(\mathbf{R}) \,
  \psi_{int}^*(\vec{r}^{(1)}_1, \cdots, \vec{r}^{(2)}_{N^{(2)}})
  \Big( \sum_{i=1}^{N^{(1)}} \frac{\mathbf{p}^{(1)2}_i}{2m^{(1)}}
  + \sum_{i=1}^{N^{(2)}} \frac{\mathbf{p}^{(2)2}_i}{2m^{(2)}} \Big)
  \psi_{int}(\vec{r}^{(1)}_1, \cdots, \vec{r}^{(2)}_{N^{(2)}})
\nonumber\\
&&+\frac{1}{2} \int \! d\vec{r} \, d\vec{r'} \, \gamma^{(1)}_{int}(\vec{r},\vec{r'}) \, u^{(1)}(\vec{r}-\vec{r'}) 
+\frac{1}{2} \int \! d\vec{r} \, d\vec{r'} \, \gamma^{(2)}_{int}(\vec{r},\vec{r'}) \, u^{(2)}(\vec{r}-\vec{r'}) 
\nonumber\\
&&+\int \! d\vec{r} \, d\vec{r'} \, \gamma^{(12)}_{int}(\vec{r},\vec{r'}) \, u^{(12)}(\vec{r}-\vec{r'}) 
\nonumber\\
&&+\int d\vec{r} v^{(1)}_{int}(\vec{r}) \rho^{(1)}_{int}(\vec{r})
+\int d\vec{r} v^{(2)}_{int}(\vec{r}) \rho^{(2)}_{int}(\vec{r})
\, .
\end{eqnarray}
\end{widetext}
To recover the "internal" KS scheme, we assume, as in the
traditional KS scheme, that there exist, \textit{in the c.m.\ frame} (the $\rho_{int}^{(l)}$ being defined in the c.m. frame, see \S \ref{par:def}), 
two non-interacting systems
(i.e. two local single-particle potentials $v^{(l)}_S$):
\begin{eqnarray}
\label{eq:KS}
\lefteqn{\Big( \frac{\vec{p}^2}{2m^{(l)}} + v^{(l)}_S(\mathbf{r}) \Big)\varphi^{(l)i}_{int}(\mathbf{r}) = \epsilon^{(l)}_i \varphi^{(l)i}_{int}(\mathbf{r}),}
\nonumber\\
&&\hspace{30mm} l=1,2 \quad , \quad 
i = 1\dots N^{(l)}
,
\end{eqnarray}
which reproduce \textit{exactly} the densities $\rho^{(l)}_{int}$ of the interacting system:
\begin{eqnarray}
\rho^{(l)}_{int}(\mathbf{r}) =\sum_{i=1}^{N^{(l)}} |\varphi^{(l)i}_{int}(\mathbf{r})|^2
.
\label{eq:rho_int}
\end{eqnarray}
Note that, even if only ($N^{(1)}+N^{(2)}-1$) coordinates are sufficient to describe internal properties, 
we introduce in the KS scheme ($N^{(1)}+N^{(2)}$) orbitals, i.e. the same number that the number of particles. 
\JMcomm{
The reason is discussed in \S~\ref{par:ks_orb}.
}

In Eq.~(\ref{eq:KS}) we implicitly supposed that the particles are Fermions \JMcomm{(i.e. that the $\{\varphi^{(l)i}_{int}\}$ are orthonormal)}.
A KS scheme to describe Boson condensates can be defined similarly by choosing identical $\varphi^{(l)i}_{int}$ for a given kind of particles $(l)$.
Uniqueness of the potential $v^{(l)}_S(\mathbf{r})$ for a given density $\rho^{(l)}_{int}(\mathbf{r})$
is ensured by a direct application of traditional DFT formalism to each KS equation (\ref{eq:KS}).
Of course, the question of the validity of the KS hypothesis, known as the \textit{non-interacting v-representability} problem,
remains, at least qualitatively, the same way that in traditional DFT \cite{Dre90} (but can quantitatively be different).

To use notation similar to the traditional notation, we add and substract from the internal energy (\ref{eq:newE2}) the internal Hartree energies
$
E^{(l)}_{H}[\rho^{(l)}_{int}] = \frac{1}{2} \int \! d\vec{r} \, d\vec{r'} \, 
\rho^{(l)}_{int}(\vec{r}) \, \rho^{(l)}_{int}(\vec{r'}) \, u^{(l)}(\vec{r}-\vec{r'})
$,\\
the "coupling" Hartree term\\
$E^{(12)}_{H}[\rho^{(1)}_{int},\rho^{(2)}_{int}]=\int d\vec{r} d\vec{r'} \rho^{(1)}_{int}(\vec{r}) \rho^{(2)}_{int}(\vec{r'}) u^{(12)}(\vec{r}-\vec{r'})$
and the non-interacting kinetic energy terms
$\sum_{i=1}^{N^{(l)}} (\varphi^{(l)i}_{int}|\frac{\vec{p}^2}{2m^{(l)}}|\varphi^{(l)i}_{int})$.
This permits us to rewrite the internal energy (\ref{eq:newE2}) as
\begin{eqnarray}
\lefteqn{E_{int}[{\rho}^{(1)}_{int},{\rho}^{(2)}_{int}]= }
\nonumber\\
&&\sum_{i=1}^{N^{(1)}} (\varphi^{(1)i}_{int}|\frac{\vec{p}^2}{2m^{(1)}}|\varphi^{(1)i}_{int})
+ \sum_{i=1}^{N^{(2)}} (\varphi^{(2)i}_{int}|\frac{\vec{p}^2}{2m^{(2)}}|\varphi^{(2)i}_{int}) 
\nonumber\\
&& + E^{(1)}_{H}[\rho^{(1)}_{int}] + E^{(2)}_{H}[\rho^{(2)}_{int}] 
+ E^{(12)}_{H}[\rho^{(1)}_{int},\rho^{(2)}_{int}]
\nonumber\\
&&
+ E^{(1)}_{XC}[\rho^{(1)}_{int},\rho^{(2)}_{int}] + E^{(2)}_{XC}[\rho^{(1)}_{int},\rho^{(2)}_{int}] + E^{(12)}_{C}[\rho^{(1)}_{int},\rho^{(2)}_{int}] 
\nonumber\\
&&+\int d\vec{r} v^{(1)}_{int}(\vec{r}) \rho^{(1)}_{int}(\vec{r})
+\int d\vec{r} v^{(2)}_{int}(\vec{r}) \rho^{(2)}_{int}(\vec{r})
.
\label{eq:action4}
\end{eqnarray}

The internal exchange-correlation energy for the identical particles of kind $(l)$ is defined as
\begin{widetext}
\begin{eqnarray}
\label{eq:Axc}
E^{(l)}_{XC}[\rho^{(1)}_{int},\rho^{(2)}_{int}]
&=& \frac{1}{2} \int \! d\vec{r} \, d\vec{r'} \, 
      \Big[ \gamma^{(l)}_{int}(\vec{r},\vec{r'}) - \rho^{(l)}_{int}(\vec{r}) \, \rho^{(l)}_{int}(\vec{r'}) \Big] \, 
      u^{(l)}(\vec{r}-\vec{r'})
\\
&&    + \sum_{i=1}^{N^{(l)}} \Big[
      \int \! d\vec{r}^{(1)}_1 \cdots d\vec{r}^{(2)}_{N^{(2)}} \delta(\mathbf{R}) \,
      \psi_{int}^*(\vec{r}^{(1)}_1 \cdots \vec{r}^{(2)}_{N^{(2)}})
      \frac{\mathbf{p}^{(l)2}_i}{2m^{(l)}}
      \psi_{int}(\vec{r}^{(1)}_1 \cdots \vec{r}^{(2)}_{N^{(2)}})
      - (\varphi^{(l)i}_{int}|\frac{\vec{p}^{2}}{2m^{(l)}}|\varphi^{(l)i}_{int}) \Big]
\nonumber
.
\end{eqnarray}
\end{widetext}
Note that, since the KS assumption implies $\varphi^{(l)i}_{int}[\rho^{(l)}_{int}]$ \cite{Dre90},
$E^{(l)}_{XC}$ can well be written as a functional of $\rho^{(1)}_{int}$ and $\rho^{(2)}_{int}$.

We see that $E^{(l)}_{XC}$ contains the exchange-correlations that come from the interaction $u^{(l)}$ (first line of (\ref{eq:Axc})),
but also the correlations contained in the interacting kinetic energy (second line of (\ref{eq:Axc})).
Concerning these correlations, it is clear that they
come, on the one hand, from the correlations neglected in the 
traditional independent-particle framework, but also from the c.m.\ 
correlations (the $\delta(\mathbf{R})$ term in the previous expression).
The kinetic energy term is the only one that \textit{explicitly} contains those correlations
because they directly affect the motions of the particles in the c.m. frame.
Indeed, in this frame, if one particle moves in one direction,
the other particles will tend to move in the opposite direction.
As quantum particles are described by wave functions, they always have associated zero-point motions.
Those zero-point motions are coupled by the c.m.\ correlations, which is specific to quantum systems;
the corresponding quantum part of the c.m. correlations is included in $E^{(l)}_{XC}$
\footnote{
\label{par:foot1}
We take the opportunity to underline a key difference between the classical and quantum separation of the c.m. motion.
In classical mechanics, as particles are pointlike, this separation is done simply by a coordinate change. 
It can be done using Jacobi coordinates (the redundant coordinate is then treated explicitly)
or the c.m. frame coordinates (which lead to a similar final result because the redundant coordinate is 
implicitly taken into account through the translational symmetry of the Hamiltonian).
In quantum mechanics, particles are described by wave functions. Then, 
one has to separate the non normalizable part of the laboratory wave function, which can be done using Jacobi coordinates.
Another consequence is that, \textit{in the c.m. frame},
all points of space that satisfy $m^{(1)} \sum_{i=1}^{N^{(1)}}\vec{r}^{(1)}_j + m^{(2)} \sum_{i=1}^{N^{(2)}}\vec{r}^{(2)}_i=0$ 
(and only those points) are allowed.
This couples the zero-point motions and produces a purely quantum contribution
to the c.m. correlations.
}.

The inclusion of the c.m. correlations in the exchange-correlation functional is 
the major difference with traditional DFT \cite{Hoh64,Koh65}
and previously developped multicomponent DFT formalisms \cite{Cap82,Kre01,Kre08,Gid98,Cha08,Cha09},
and opens the way to the search for a local c.m. correlations potential,
which would \textit{a priori} be computationally much less costly than the projection techniques used, for instance, in nuclear physics
\cite{Schm01a,Schm04,Pei62,Ben03}.
Note also that the inclusion of the c.m. correlations explains why $E^{(l)}_{XC}$ is a functional of both $\rho^{(1)}_{int}$ and $\rho^{(2)}_{int}$, 
and not only of $\rho^{(l)}_{int}$ as one might have expected:
those correlations necessarilly couple the particles $(1)$ and $(2)$, thus their densities
(this will be underlined from another point of view in \S \ref{par:link1}).

We emphasize the fact that a part of the total c.m. correlations is contained in $E^{(1)}_{XC}$ and another part in $E^{(2)}_{XC}$.
Indeed, it is the interacting kinetic energy term as a whole that contains explicitly the c.m. correlations,
and this term is split in both $E^{(1)}_{XC}$ 
and $E^{(2)}_{XC}$.
The decomposition we choose for the energy functional is of course not unique.
But even if it does not include all the c.m. correlations in one specific functional,
its advantage is that the meaning of the obtained functionals is clear and that
it permits us to recover the functional forms of "one kind of particle" traditional and internal DFT
in the limits that will be discussed in \S \ref{par:tradDFT}.

The internal coupling correlation energy between particles of kinds $(1)$ and $(2)$ is defined as
\begin{eqnarray}
\label{eq:Ecoupling}
\lefteqn{E^{(12)}_{C}[\rho^{(1)}_{int},\rho^{(2)}_{int}]= }
\\
&& \int \! d\vec{r} \, d\vec{r'} \, 
      \Big[ \gamma^{(12)}_{int}(\vec{r},\vec{r'}) - \rho^{(1)}_{int}(\vec{r}) \, \rho^{(2)}_{int}(\vec{r'}) \Big] \, 
      u^{(12)}(\vec{r}-\vec{r'})
\nonumber
.
\end{eqnarray}
There is of course no exchange energy in (\ref{eq:Ecoupling}) because the particles $(1)$ and $(2)$ are not identical.
As this term does not contain \textit{explicitly} the c.m. correlations, the spirit of the already developped approximations found in \cite{Cha08,Kre01,Kre08}
remains suitable within the formalism presented here, at least for molecular systems, especially that of \cite{Cha08} which is directly applicable
to a "coupling" correlation energy written in the form (\ref{eq:Ecoupling}).

The remaining task is to minimize the internal energy (\ref{eq:action4}) 
so as obtain the equations of motion (which define $\rho_{int}^{(l)}$).
Varying $E_{int}[\rho^{(1)}_{int},\rho^{(2)}_{int}]$ with respect to $\varphi^{(l)i*}_{int}$, and imposing 
\JMcomm{(ortho-)}normality of the $\{\varphi^{(l)i}_{int}\}$,
\begin{eqnarray}
\frac{\delta}{\delta \varphi^{(l)i*}_{int} (\vec{r})}
\Big(   E_{int}[\rho^{(1)}_{int},\rho^{(2)}_{int}] 
      - \sum_{m=1}^{2} \sum_{i=1}^{N^{(m)}} \epsilon^{(m)}_i \, (\varphi^{(m)i}_{int}|\varphi^{(m)i}_{int}) 
\Big)
=0
\nonumber
\end{eqnarray}
leads to "internal" Kohn-Sham equations (\ref{eq:KS}) for the $\{\varphi^{(l)i}_{int}\}$ with
\begin{widetext}
\begin{eqnarray}
\label{eq:varphi_i}
v^{(l)}_S(\mathbf{r}) &=&
\int d\vec{r'} \rho^{(l)}_{int}(\vec{r'}) u^{(l)}(\vec{r}-\vec{r'}) + U^{(l_{/l})}_{XC}[\rho^{(1)}_{int},\rho^{(2)}_{int}](\vec{r}) + v^{(l)}_{int}(\vec{r})
\nonumber\\
&&
+ \int d\vec{r'} \rho^{(m)}_{int}(\vec{r'}) u^{(12)}(\vec{r}-\vec{r'}) + U^{(m_{/l})}_{C}[\rho^{(1)}_{int},\rho^{(2)}_{int}] (\vec{r})
+ U^{(12_{/l})}_{C}[\rho^{(1)}_{int},\rho^{(2)}_{int}] (\vec{r})
\quad\quad\quad , \quad for \hspace{1mm} m\ne l
.
\end{eqnarray}
\end{widetext}
We keep $v^{(l)}_{int}$ for generality but recall that it can be set to zero in the case of an isolated self-bound system.
In addition to the classical parts of the interaction (the Hartree potentials for the $(l)-(l)$ and $(1)-(2)$ interactions,
and the optional external potentials) appear purely quantum
potentials, which are defined as follows.

The exchange-correlation potential for the particles of kind $(l)$ is
\begin{eqnarray}
U^{(l_{/l})}_{XC}[\rho^{(1)}_{int},\rho^{(2)}_{int}] (\vec{r}) 
= \frac{\delta E^{(l)}_{XC}[\rho^{(1)}_{int},\rho^{(2)}_{int}]}{\delta \rho^{(l)}_{int}(\vec{r})}
.
\end{eqnarray}
It is functional of $\rho^{(1)}_{int}$ and $\rho^{(2)}_{int}$, and not only of $\rho^{(l)}_{int}$,
for the reasons discussed above and linked to the c.m. correlations.
This potential contains the "standard" exchange-correlation for the $(l)-(l)$ interaction 
and the part of the c.m. correlations that is contained in $E^{(l)}_{XC}$.

The other part of the c.m. correlations is included in $E^{(m\ne l)}_{XC}$ (because of the decomposition we choose).
Thus there also appear
complementary pure c.m. correlations potentials
\begin{eqnarray}
U^{(m_{/l})}_{C}[\rho^{(1)}_{int},\rho^{(2)}_{int}] (\vec{r}) 
= \frac{\delta E^{(m)}_{XC}[\rho^{(1)}_{int},\rho^{(2)}_{int}]}{\delta \rho^{(l)}_{int}(\vec{r})} 
\quad , \quad for \hspace{1mm}m\ne l
\nonumber.
\end{eqnarray}

Finally, the coupling correlation potential is
\begin{eqnarray}
U^{(12_{/l})}_{C}[\rho^{(1)}_{int},\rho^{(2)}_{int}] (\vec{r}) 
= \frac{\delta E^{(12)}_{C}[\rho^{(1)}_{int},\rho^{(2)}_{int}]}{\delta \rho^{(l)}_{int}(\vec{r})}
.
\label{eq:Uxc}
\end{eqnarray}
%

All those potentials are local as expected, which is the computational power of DFT.
One can note that Eqs.~(\ref{eq:KS}) and (\ref{eq:varphi_i}) are symmetrical under the exchange $(1)$ $\leftrightarrow$ $(2)$.

Note that the first line of Eq.~(\ref{eq:varphi_i}) contains the terms that describe the interaction between the identical particles of kind $(l)$,
which are close to those found in the "one kind of particle" KS scheme \cite{Koh65}.
The second line of Eq.~(\ref{eq:varphi_i}) contains terms that are specific to the KS equations for two kinds of particles.
They describe the interaction of the particles of kind $(l)$ with the particles of kind $(m\ne l)$.

\JMcomm{
\subsection{
Why do we introduce a number of orbitals equal to the number of particles in the KS scheme?
}
\label{par:ks_orb}
The internal DFT formalism raises the following question (we consider only one kind of fermion in this section to simplify the discussion):
%
since, for a self-bound system composed of $N$ particles, only ($N-1$) coordinates are sufficient to describe internal properties,
why do we introduce $N$ orbitals in the internal KS scheme?
We mention that the question of the number of orbitals to introduce in the KS scheme also occurs in traditional DFT. 
Indeed, this number is not imposed by the theory; it 
has been fixed "by hand" to $N$ orthonormal orbitals
(introducing an auxiliary system of $N$ non-interacting particles).
From now, nothing would fundamentally forbid us from introducing another number 
of orbitals (eventually non-orthogonal with effective masses).
This would of course change the form of the KS energy functional,
which will "adapt" to the change of the number of orbitals.
More precisely, since to a choosen number $M$ of orbitals corresponds a specific definition of the non-interacting "kinetic energy",
it would change the part of $E_{XC}[\rho]$ that corresponds to the difference between the interacting 
kinetic energy and the non-interacting "kinetic energy"
(i.e. the second line of (\ref{eq:Axc})), 
that we note $E_{\Delta kin}[\rho]$.
For instance, in the case where only one KS orbital would be introduced, 
$E^{(M=1)}_{\Delta kin}[\rho]$ would be drastically different from
$E^{(M=N)}_{\Delta kin}[\rho]$, but 
it is easy to show that the corresponding single KS equation would lead to the exact interacting ground-state density,
i.e. to the same result that the one obtained by direct variation of the HK functional by $\rho$,
so that the non-interacting v-representability is perfectly achieved
(since the KS non-interacting "kinetic energy" 
can then trivially be written as an explicit functional of $\rho$ ($\varphi_1=\sqrt{\rho}$) and thus its variation is defined for 
all densities for which $\sqrt{\rho}$ is differentiable, including the interacting v-representable ones).}

\JMcomm{\textit{Purely formally speaking},
the question of the optimum number of orbitals 
to introduce in the KS scheme 
can be reformulated as follows:
is the non-interacting v-representability better achieved with a certain number of orbitals $M$ ($>1$)?
This question is still open in traditional (and internal) DFT.
}

\JMcomm{
\textit{But practically speaking}, many advantages justify the introduction of $N$ orbitals:
\begin{enumerate}[(i)]
\item 
Even if only ($N-1$) coordinates are sufficient to describe the internal properties of a self-bound system,
they still describe a system of $N$ particles.
We thus have to introduce $N$ orthonormal orbitals 
if we want them to be interpreted (to first order only) as single-particle orbitals
and obtain a scheme comparable
to actual mean-field like calculations with effective interactions.
In particular, the highest KS eigenvalue $\epsilon_N$
is then the negative of the ionization (or the separation) energy with the exact functional \cite{Per82,Per97}.
\item
The anti-symmetrization is explicit in terms of the $N$ particles coordinates only.
Introduction of $N$ orbitals permits us to explicitly take into account the anti-symmetrization in the definition of $\rho$,
while the anti-symmetrization would be implicit if another number of orbitals was introduced
and thus more difficult to obtain because it would have to be more precisely taken into account in parametrizations of $E_{XC}[\rho]$.
%
\item
It leads straightforwardly to the classical (point-like) limit, see appendix \ref{app:jacobi5}.
\item
It \textit{a priori} permits one to obtain, in the general case, 
the non-interacting kinetic energy that is the closest to the interacting kinetic energy (i.e. the smallest $E^{(M)}_{\Delta kin}[\rho]$).
Indeed, $E^{(M=N)}_{\Delta kin}[\rho]$ contains only the correlations energy part of the interacting kinetic energy,
which is generally a correction.
It is 
essentially for quite small self-bound systems and within the internal DFT formalism
that $E^{(M=N)}_{\Delta kin}[\rho]$ can be more than a correction,
because it contains the c.m. correlations, which can be large for those systems.
Then, one track is to conserve $N$ orbitals and to develop precise parametrizations for $E^{(M=N)}_{\Delta kin}[\rho]$ 
(which is not achieved by actual functionals);
this would permit us to preserve the advantages listed above.
But the previous discussion also let the door open to another possible track: study whether the introduction of a number of orbitals $M\ne N$ 
(for instance $M=N-1$ with effective masses) would permit us to 
obtain a smaller $E^{(M)}_{\Delta kin}[\rho]$,
so that difficulties in parametrizing $E^{(M)}_{\Delta kin}[\rho]$ would affect the result as little as possible.
This goes beyond the scope of this paper.
\end{enumerate}
}

\subsection{Link with previous multicomponent DFT formalisms.}
\label{par:link1}

\JMcomm{The previous multicomponent DFT formalisms found in Refs.~\cite{Cap82,Gid98,
Cha08,Cha09} have been developped without separation of the c.m. motion,
as already underlined in Ref.~\cite{Gid98}.
The definitions of the energy functionals were based on the \textit{laboratory} wave function $\psi$ and densities $\rho^{(l)}$.}
To understand the key differences with those previous formalisms,
we switch to the Levy-Lieb constrained search formulation \cite{Lev79,Lie83},
within which they have been developped.
The corresponding energy
can be decomposed in a form similar to Eq.~(\ref{eq:action4}), where only the definitions of $E^{(l)}_{XC}$ and $E^{(12)}_{C}$
change (i.e. the fourth line of Eq.~(\ref{eq:action4})) \cite{Gid98,Cha08,Cha09}:
%
\begin{widetext}
\begin{eqnarray}
E^{(l)}_{XC}[\rho^{(1)}_{int},\rho^{(2)}_{int}] \rightarrow&&
\label{eq:prevmDFT}\\
E^{(l)}_{XC}[\rho^{(l)}]&=&
      \min_{\psi^{(l)}\rightarrow\rho^{(l)}} (\psi^{(l)}|\hat{T}^{(l)}+\hat{U}^{(l)}|\psi^{(l)})
      - \min_{\psi^{(l)}_{SD}\rightarrow\rho^{(l)}} (\psi^{(l)}_{SD}|\hat{T}^{(l)}|\psi^{(l)}_{SD})
      -E^{(l)}_{H}[\rho^{(l)}]
,
\nonumber\\
E^{(12)}_{C}[\rho^{(1)}_{int},\rho^{(2)}_{int}] \rightarrow&&
\nonumber\\
E^{(12)}_{C}[\rho^{(1)},\rho^{(2)}]&=&
       \min_{\psi\rightarrow\rho^{(1)},\rho^{(2)}} (\psi|\hat{T}^{(l)}+\hat{T}^{(2)}+\hat{U}^{(1)}+\hat{U}^{(2)}+\hat{U}^{(12)}|\psi)
       -\min_{\psi^{(1)}\rightarrow\rho^{(1)}} (\psi^{(1)}|\hat{T}^{(1)}+\hat{U}^{(1)}|\psi^{(1)})
\nonumber\\
       &&-\min_{\psi^{(2)}\rightarrow\rho^{(2)}} (\psi^{(2)}|\hat{T}^{(2)}+\hat{U}^{(2)}|\psi^{(2)})
       - E^{(12)}_{H}[\rho^{(1)},\rho^{(2)}]
.
\nonumber
\end{eqnarray}
\end{widetext}
The $\psi$ are the interacting ($N^{(1)}+N^{(2)}$)-particles states, the $\psi^{(l)}_{SD}$ are the non-interacting $N^{(l)}$-particles states (Slater Determinants),
the $\psi^{(l)}$ are the interacting $N^{(l)}$-particles states
\JMcomm{and $\hat{T}^{(l)}=\sum_{i=1}^{N^{(l)}} \frac{{\hat{p}_i^{(l)2}}}{2m^{(l)}}$ is the kinetic energy operator for particles of kind ($l$)}. 
The interest of the introduction of the $\psi^{(l)}$ (and of the constrained search formulation)
is that it permits us to define $E^{(l)}_{XC}$ as a functional of $\rho^{(l)}$ only
(and not of both $\rho^{(1)}$ and $\rho^{(2)}$).
\JMcomm{The constrained search formulation is by definition restricted to densities and thus states that are normalizable
(the variational principle being defined only for systems for which the wave function is normalizable \cite{Lie83,Bran}).
However, as the functionals are formulated in terms of the laboratory density (i.e. the c.m. motion is not separated), the equations obtained after minimization of the energy functional should lead to full delocalization of the density.
Moreover, as the theory is not formulated in terms of internal properties, 
the c.m. correlations 
do not appear explicitly in the functional.
}
%
%
%

Internal DFT overcomes those problems 
because it is based on the internal wave function,
which is by definition always normalizable for self-bound systems, 
and is the one which defines the observables of experimental interest \cite{Mes09}.
For better comparison with the previous multicomponent DFT formalisms, we reformulate 
\JMcomm{the energies $E^{(l)}_{XC}$ and $E^{(12)}_{C}$ defined in \S \ref{par:intKS}}
within the constrained search formulation:
\begin{widetext}
\begin{eqnarray}
E^{(l)}_{XC}[\rho^{(1)}_{int},\rho^{(2)}_{int}] 
&=&
      \min_{\psi_{int}\rightarrow\rho^{(1)}_{int},\rho^{(2)}_{int}} 
      (\psi_{int}|\hat{T}^{(l)}+\hat{U}^{(l)}|\psi_{int})
      - \min_{\psi^{(l)}_{SD}\rightarrow\rho^{(l)}_{int}} (\psi^{(l)}_{SD}|\hat{T}^{(l)}|\psi^{(l)}_{SD})
      -E^{(l)}_{H}[\rho^{(l)}_{int}]
,
\label{eq:newmDFT}\\
E^{(12)}_{C}[\rho^{(1)}_{int},\rho^{(2)}_{int}] 
&=&
        \min_{\psi_{int}\rightarrow\rho^{(1)}_{int},\rho^{(2)}_{int}} 
       (\psi_{int}|\hat{T}^{(1)}+\hat{T}^{(2)}+\hat{U}^{(1)}+\hat{U}^{(2)}+\hat{U}^{(12)}|\psi_{int})
\nonumber\\
       &&-\min_{\psi_{int}\rightarrow\rho^{(1)}_{int},\rho^{(2)}_{int}}(\psi_{int}|\hat{T}^{(1)}+\hat{U}^{(1)}|\psi_{int})
         -\min_{\psi_{int}\rightarrow\rho^{(1)}_{int},\rho^{(2)}_{int}} (\psi_{int}|\hat{T}^{(2)}+\hat{U}^{(2)}|\psi_{int})
         - E^{(12)}_{H}[\rho^{(1)}_{int},\rho^{(2)}_{int}]
.
\nonumber
\end{eqnarray}
\end{widetext}
The $\psi_{int}$ are the interacting internal states which can be expressed with ($N^{(1)}+N^{(2)}-1$) Jacobi coordinates
\JMcomm{\footnote{\JMcomm{
Note that when the integrals involving $\psi_{int}$ are written in the ($N^{(1)}+N^{(2)}$) particles coordinates
representation, following similar steps as in~(\ref{eq:rel}), a $\delta(\mathbf{R})$ appears explicitly.
}}}.
The $\psi^{(l)}_{SD}$ are the non-interacting $N^{(l)}$-particles states.
%

Note that the choice we made in (\ref{eq:newmDFT}) for the $\tilde\psi$ used 
in the minimization of $(\tilde\psi|\hat{T}^{(l)}+\hat{U}^{(l)}|\tilde\psi)$ is not unique
(because there is a cancelation between $E^{(l)}_{XC}$ and $E^{(12)}_{C}$).
One could, for instance, use interacting states $\psi^{(l)}_{int}$ 
expressible with ($N^{(l)}-1$) Jacobi coordinates
which lead to $\rho^{(l)}_{int}$,
instead of the ($N^{(1)}+N^{(2)}-1$) Jacobi coordinates state
leading to $\rho^{(1)}_{int}$ \textit{and} $\rho^{(2)}_{int}$
\footnote{
\JMcomm{Since the c.m. motion is "substracted" in our formalism, the most pertinent interacting states to use 
are of course internal ones, i.e. expressible in terms of Jacobi coordinates.}
}.
Even if formally correct, this is not the choice used in (\ref{eq:newmDFT}) mainly because it would lead to interpretation
difficulties.
Indeed, it amounts to considering
that the particles of kinds $(1)$ and $(2)$ have independent 
c.m. motions~
\JMcomm{\footnote{\JMcomm{
Indeed, following similar steps as in~(\ref{eq:rel}), the integral $(\psi^{(l)}_{int}|\hat{T}^{(l)}+\hat{U}^{(l)}|\psi^{(l)}_{int})$ can be written 
in terms of the $N^{(l)}$ coordinates of particles ($l$).
Then, a $\delta(\mathbf{R}^{(l)})$ appears explicitly.
}}}, 
while the c.m. motion is common for the system as a whole.
%
Moreover, this would bring no gain because it would move the explicit inclusion of total c.m. correlations
to $E^{(12)}_{C}$,
so that previously developped approximations for this term \cite{Cha08,Kre01,Kre08} would not be fully usable,
contrary to the formulation proposed in this paper.
Another advantage of the decomposition we choose in this paper is that it permits a clear link
with "one kind of particle" standard and internal DFT, which will be discussed in \S \ref{par:tradDFT}.


We now discuss the link with the slightly different work found in \cite{Kre01,Kre08}.
First of all, this work is by its very nature essentially suited for the description of molecular systems,
since the electronic coordinates are transformed to the body-fixed frame
\textit{independently} from the nuclear coordinates 
(this "decoupling" is possible in molecular systems due to the large difference of masses between electrons and nuclei, see appendix \ref{app:jacobi3};
independent transformations are no longer possible for a self-bound system where there is no large difference of masses between the different kinds of particles constituting it).
Second, this work treats the breaking of rotational symmetry, which
goes further than the other multicomponent DFT formalisms (we don't treat rotational symmetry in this article, but we are working on the problem).
Third, the nuclear $N^{(n)}$-body density is used as a basic variable instead of the nuclear one-body density,
for easier description of collective phenomena
(but it is of course less easy to handle numerically for quite large systems).

\JMcomm{The 
difficulty discussed 
above remains: even if the electronic coordinates are transformed to the body-fixed frame,
since the c.m. wave function is not \textit{separated} from the beginning,
all the densities are defined through a 
fully delocalized
wave function (for a ground state).}
%
But numerically speaking this does not cause a real problem because:
\begin{enumerate}[(i)]
\item The KS step breaks by essence the translational symmetry,
so that it forces the densities to become \JMcomm{localized}.
\item In Refs.~\cite{Kre01,Kre08} the KS scheme is formulated in terms of the $N^{(n)}$-body nuclear density and,
in the presented practical calculations,
the nuclear c.m. has been separated at a second step, leading to the inclusion of 
the c.m. correlations in the corresponding $N^{(n)}$-body auxiliary nuclear wave function.
\end{enumerate}

In the formalism presented in the present article, the c.m. motion has been separated from the beginning, 
leading to a proper formulation 
of the HK theorem \JMcomm{in terms of the internal densities} 
and the explicit inclusion of the c.m. correlations in the functional. 
This formalism is suited to study all self-bound systems.

We now study thoroughly the application to molecular systems, which permits us to make explicit the link between this formalism and traditional DFT,
and then detail the application to atomic nuclei and mixtures of $^3$He and $^4$He droplets.

\section{Application to molecular systems and limit of traditional DFT.}
\label{par:tradDFT}

Electrons (we note $l=e$) are Fermions, so that exchange and correlations are included in the functional $E^{(e)}_{XC}$.
Nuclei (we note $l=n$) are approximated as point-like and can be Fermions or Bosons (in the latter case, the functional $E^{(n)}_{XC}$ contains only correlations and all the $\varphi^{(n)i}_{int}$ are identical).
The internal potentials $v^{(l)}_{int}$ can then describe internal effects of a polarization potential (i.e. a voltage applied to the system), 
for instance in the stationary case,
or internal effects of lasers used in molecular irradiation experiments, for instance in the time-dependent case~\cite{Mes09-2}.

\subsection{Simplification of the formalism due to the large difference of masses.}
\label{par:tradDFT1}

In \JMcomm{molecular systems}, the nuclei are much heavier than the electrons, i.e. $m^{(n)}>>m^{(e)}$.
It thus should be a very good approximation to consider $\mathbf{R}=\mathbf{R}^{(n)}$
and apply Jacobi coordinates to the nuclear coordinates only,
so that the nuclei will be described by ($N^{(n)}-1$) Jacobi coordinates and will carry all the c.m. correlations.
As a consequence, the electrons are not concerned by the redundant coordinates problem (i.e. the c.m. correlations)
and remain described by $N^{(e)}$ coordinates in the frame attached to the c.m. of the nuclei.
See appendix \ref{app:jacobi3} for better understanding of this mechanism from the point of view of Jacobi coordinates
and appendix \ref{app:jacobi4} from the point of view of the functionals.

As a direct consequence, the electronic exchange-correlation functional $E^{(e)}_{XC}$ no longer contains explicitly the c.m.\ 
correlations and its form becomes comparable to that of the exchange-correlation functional of "one kind of particle" traditional DFT \cite{Dre90},
see Eq. (\ref{eq:Exc_e}). All the c.m.\ correlations are explicitly included in the 
nuclear exchange-correlation functional $E^{(n)}_{XC}$ only and its form becomes comparable to the form of the exchange-correlation functional 
of "one kind of particle" internal DFT \cite{Mes09}, see Eq. (\ref{eq:Exc_n}).
Since it is only the c.m. coupling between the particles $(l)$ and $(m\ne l)$ that produces a $\rho^{(m \ne l)}_{int}$ dependence in
$E^{(l)}_{XC}$ (see discussion following Eq.~(\ref{eq:Axc})), we have:
\begin{eqnarray}
\frac{\delta E^{(l)}_{XC}[\rho^{(1)}_{int},\rho^{(2)}_{int}]}{\delta \rho^{(m)}_{int}(\vec{r})} = 0
\quad , \quad for \hspace{1mm}m\ne l \quad \Rightarrow \quad E^{(l)}_{XC}[\rho^{(l)}_{int}]
\nonumber
.
\end{eqnarray}
The internal energy (\ref{eq:action4}) thus becomes
\begin{eqnarray}
\lefteqn{E_{int}[{\rho}^{(n)}_{int},{\rho}^{(e)}_{int}] =}
\nonumber\\
&& \sum_{i=1}^{N^{(n)}} (\varphi^{(n)i}_{int}|\frac{\vec{p}^2}{2m^{(n)}}|\varphi^{(n)i}_{int})
+ \sum_{i=1}^{N^{(e)}} (\varphi^{(e)i}_{int}|\frac{\vec{p}^2}{2m^{(e)}}|\varphi^{(e)i}_{int}) 
\nonumber\\
&& + E^{(n)}_{H}[\rho^{(n)}_{int}] + E^{(e)}_{H}[\rho^{(e)}_{int}] 
+ E^{(ne)}_{H}[\rho^{(n)}_{int},\rho^{(e)}_{int}]
\nonumber\\
&&
+ E^{(n)}_{XC}[\rho^{(n)}_{int}] + E^{(e)}_{XC}[\rho^{(e)}_{int}] + E^{(ne)}_{C}[\rho^{(n)}_{int},\rho^{(e)}_{int}] 
\nonumber\\
&&+\int d\vec{r} v^{(n)}_{int}(\vec{r}) \rho^{(n)}_{int}(\vec{r})
+\int d\vec{r} v^{(e)}_{int}(\vec{r}) \rho^{(e)}_{int}(\vec{r})
,
\label{eq:action5}
\end{eqnarray}
and the KS potentials (\ref{eq:varphi_i}) for the electrons and nuclei become
\begin{widetext}
\begin{eqnarray}
\label{eq:varphi_i2}
v^{(e)}_S(\mathbf{r}) &=&
\int d\vec{r'} \rho^{(e)}_{int}(\vec{r'}) u^{(e)}(\vec{r}-\vec{r'}) + U^{(e_{/e})}_{XC}[\rho^{(e)}_{int}](\vec{r}) + v^{(e)}_{int}(\vec{r})
+ \int d\vec{r'} \rho^{(n)}_{int}(\vec{r'}) u^{(en)}(\vec{r}-\vec{r'}) + U^{(en_{/e})}_{C}[\rho^{(e)}_{int},\rho^{(n)}_{int}] (\vec{r})
,
\nonumber\\
v^{(n)}_S(\mathbf{r}) &=&
\int d\vec{r'} \rho^{(n)}_{int}(\vec{r'}) u^{(n)}(\vec{r}-\vec{r'}) + U^{(n_{/n})}_{XC}[\rho^{(n)}_{int}](\vec{r}) + v^{(n)}_{int}(\vec{r})
+ \int d\vec{r'} \rho^{(e)}_{int}(\vec{r'}) u^{(en)}(\vec{r}-\vec{r'}) + U^{(en_{/n})}_{C}[\rho^{(e)}_{int},\rho^{(n)}_{int}] (\vec{r})
.
\nonumber\\
\end{eqnarray}
\end{widetext}

Since $E^{(e)}_{XC}[\rho^{(e)}_{int}]$ is comparable to the exchange-correlation functional of traditional DFT,
standard approximations remain pertinent, for example the widely used Local Density Approximation (LDA)~
\footnote{
I.e. maximum delocalization approximation (Fermi gas) and replacement, in the obtained functional, of the Fermi gas density by the true density of the system.
},
see e.g. \cite{Jon89aR}, or its extension to the Generalized Gradient Approximation (GGA) \cite{Per96a}.

It remains an open problem to find satisfying numerically manageable approximations for $E^{(n)}_{XC}[\rho^{(n)}_{int}]$,
which among others would contains the c.m. correlations energy.
It would permit us to study molecular systems where the quantum nature of the nuclei plays an important role
(for example, small molecules \cite{Gro95} and solid hydrogen \cite{Sur93}).
The LDA could not be used \textit{a priori} to approximate it
because the nuclei are generally localized. It would probably be more pertinent to start from an approximation which exploits localization
(and not delocalization as LDA does).
A crude approximation would be to do the
maximum localization approximation 
and replace, in the obtained functional, 
the corresponding classical density by the true density of the system.
Appendix \ref{app:jacobi5} details how to make properly the classical (pointlike) approximation
for self-bound systems, which is not completely trivial because of the translational symmetry.
Then, the "kinetic energy correlations" part of $E_{XC}^{(n)}$
(which contains explicitly the c.m. correlations)
desappears and\\
$\gamma^{(n)}_{int}(\mathbf{r},\mathbf{r}') - \rho^{(n)}_{int}(\mathbf{r})\rho^{(n)}_{int}(\mathbf{r}')
\rightarrow \sum_{i=1}^{N^{(n)}} \rho^{(n)i}_{int} (\mathbf{r}) \rho^{(n)i}_{int} (\mathbf{r}')$,
where $\rho^{(n)i}_{int}=|\varphi^{(n)i}_{int}|^2$, see appendix \ref{app:jacobi5}.
We thus obtain $E^{(n)}_{XC}= - \sum_{i=1}^{N^{(n)}} E^{(n)}_{H} [\rho^{(n)i}_{int}]$,
which corresponds to the self-interaction correction to the ($n$)-($n$) Hartree energy,
as noticed in Refs.~\cite{Gid98,Gro96} for the standard case,
and leads to the potential $U^{(n_{/n})}_{XC}=-\int d\vec{r'} \rho^{(n)i}_{int}(\vec{r'}) u^{(n)}(\vec{r}-\vec{r'})$ 
when applied to the state $\varphi^{(n)i}_{int}$.
This approximation gives a single densities dependence and not an explicit total density dependence
(this dependence is nevertheless implicit as soon as a local KS potential is imposed, 
which is the aim of the Optimized Effective Potential method \cite{Sha53a,Tal76,Kue07aR}, 
because it implies $\varphi^{(n)i}_{int}[\rho^{(n)}_{int}]$).
This necessitates extra (numerically more costly) caution to ensure the orthonormality of the $\{\varphi^{(n)i}_{int}\}$,
see for example Refs. \cite{Per81,Ped84,Mes09-3}.

Even if this "maximally localized" approximation should be used \textit{a minima} in the absence of other approximations, 
its range of validity is not clear.
Indeed, the width of the nuclear single densities is not systematically very small \cite{Gid98,Tho69}. 
Moreover, this approximation does not permit us to evaluate the c.m. correlations contained in $E^{(n)}_{XC}$
(because they disapear in the classical limit,
see appendix \ref{app:jacobi5} and footnote \ref{par:foot1}).
Thus, the question of a suitable approximation for $E^{(n)}_{XC}$ as a functional of the one-body total nuclear density, including the c.m. correlations,
remains open.
A possible approach would be to use the semi-classical approximation \cite{Bra08} instead of the crude classical (point-like) approximation.

Concerning $E^{(ne)}_{C}[\rho^{(n)}_{int},\rho^{(e)}_{int}]$, we recall that the spirit of the 
already developped approximations found in Refs.~\cite{Cha08,Kre01,Kre08}
remain suitable within the formalism presented here, especially that of Ref.~\cite{Cha08},
but futher developments are certainly desirable.

Finally, one can note that the KS equations (\ref{eq:KS}) and (\ref{eq:varphi_i2}) still remain symmetric under the exchange $(e)$ $\leftrightarrow$ $(n)$,
but the forms of $E^{(n)}_{XC}$ and $E^{(e)}_{XC}$ (thus of $U^{(n_{/n})}_{XC}$ and $U^{(e_{/e})}_{XC}$) are different:
one contains explicitly the c.m. correlations whereas the other one does not.

We now make further approximations that permit us to recover traditional DFT and thus to understand better the 
approximations inherent to this formalism and how to eventually improve its results.

\subsection{Further approximations that enable us to recover traditional DFT.}

\subsubsection{Electron-nuclear correlation energy neglected.}
\label{par:tradDFT2}

We suppose that, in the total internal energy (\ref{eq:action5}), the electron-nuclear correlation energy $E^{(en)}_{XC}$
is negligible compared to the other energies.
This in particular implies that $|E^{(ne)}_{C}|$ is much smaller than $|E^{(n)}_{XC}|$. 
It is likely that this is a fair approximation for a certain class of molecular systems, such as:
\begin{enumerate}[(i)]
\item $E^{(n)}_{XC}$ should at least contain the ($n$)-($n$) Hartree self interaction (as discussed above),
\item electrons and nuclei are different kinds of particles,
\item nuclei are generally localized (thus not too far from classical particles).
\end{enumerate}
However, the range of validity of this approximation is not presently perfectly clear.
Indeed, there exist molecular systems for which the electrons-nuclei correlations are important 
for the description of some physical properties \cite{Cha08}. Then, one has to remain at the previous step.

Within this approximation, the total internal energy becomes
\begin{eqnarray}
\lefteqn{E_{int}[{\rho}^{(n)}_{int},{\rho}^{(e)}_{int}]= } 
\nonumber\\
&&\sum_{i=1}^{N^{(n)}} (\varphi^{(n)i}_{int}|\frac{\vec{p}^2}{2m^{(n)}}|\varphi^{(n)i}_{int})
+ \sum_{i=1}^{N^{(e)}} (\varphi^{(e)i}_{int}|\frac{\vec{p}^2}{2m^{(e)}}|\varphi^{(e)i}_{int}) 
\nonumber\\
&& + E^{(n)}_{H}[\rho^{(n)}_{int}] + E^{(e)}_{H}[\rho^{(e)}_{int}] 
+ E^{(ne)}_{H}[\rho^{(n)}_{int},\rho^{(e)}_{int}]
\nonumber\\
&&
+ E^{(n)}_{XC}[\rho^{(n)}_{int}] + E^{(e)}_{XC}[\rho^{(e)}_{int}] 
\nonumber\\
&&+\int d\vec{r} v^{(n)}_{int}(\vec{r}) \rho^{(n)}_{int}(\vec{r})
+\int d\vec{r} v^{(e)}_{int}(\vec{r}) \rho^{(e)}_{int}(\vec{r})
,
\label{eq:action7}
\end{eqnarray}
and the KS potentials (\ref{eq:varphi_i2}) become
\begin{eqnarray}
v^{(e)}_S(\mathbf{r}) &=&
\int d\vec{r'} \rho^{(e)}_{int}(\vec{r'}) u^{(e)}(\vec{r}-\vec{r'}) + U^{(e_{/e})}_{XC}[\rho^{(e)}_{int}](\vec{r}) 
\nonumber\\
&&+ v^{(e)}_{int}(\vec{r})
+ \int d\vec{r'} \rho^{(n)}_{int}(\vec{r'}) u^{(en)}(\vec{r}-\vec{r'})
,
\label{eq:varphi_i3}
\\
v^{(n)}_S(\mathbf{r}) &=&
\int d\vec{r'} \rho^{(n)}_{int}(\vec{r'}) u^{(n)}(\vec{r}-\vec{r'}) + U^{(n_{/n})}_{XC}[\rho^{(n)}_{int}](\vec{r}) 
\nonumber\\
&&+ v^{(n)}_{int}(\vec{r})
+ \int d\vec{r'} \rho^{(e)}_{int}(\vec{r'}) u^{(en)}(\vec{r}-\vec{r'}) 
\nonumber.
\end{eqnarray}
%
%
We see that, for the electrons, we have recovered the traditional "one kind of particule" KS potential
(as $U^{(e_{/e})}_{XC}$ does not contain the c.m. correlations),
with, in standard notation \cite{Hoh64,Koh65,Dre90,Koh99,DFTLN}
\begin{eqnarray}
\label{eq:vext}
v^{ext}(\vec{r})=v^{(e)}_{int}(\vec{r})+\int d\vec{r'} \rho^{(n)}_{int}(\vec{r'}) u^{(en)}(\vec{r}-\vec{r'})
.
\end{eqnarray}
The potential $v^{ext}$ contains the arbitrary one-body internal potential $v^{(e)}_{int}$
and the Hartree part of the electron-nuclear interaction (which is a functional of the nuclear one-body density $\rho^{(n)}_{int}$).
$v^{ext}$ is thus functional of $v^{(e)}_{int}$ and $\rho^{(n)}_{int}$,
which is allowed by traditional DFT
(traditional DFT is valid for any $v^{ext}$ that is not a functional of the electronic wave function, \JMcomm{thus density}, and that leads to a bound state).
The potential $v^{ext}$, which is \textit{internal} for (self-bound) molecular systems, becomes \textit{external} for the pure electronic problem.
This permits us to understand why traditional DFT is particularly well suited for the description of the electrons (only) in a molecular system,
in the frame attached to the c.m. of the nuclei.

Note that, contrary to what is sometimes \JMcomm{thought},
traditional DFT does not,
strictly speaking, involve the clamped nuclei approximation (and thus not the Born-Oppenheimer approximation either).
Indeed, traditional KS equations (from which follow the $\varphi^{(e)i}_{int}$) do not necessarily depend parametrically on the nuclear positions.
One is allowed to use an external potential of the form (\ref{eq:vext}),
so that the nuclei are described with a spatial width (i.e. quantum-mechanically),
and to specify another equation that defines $\rho^{(n)}_{int}$.
It is when this last equation is not specified (for instance if we simply impose that each nuclei is represented by a gaussian of fixed width) 
or when the nuclei are treated classically (see next section)
that the total molecular energy should be calculated for various nuclear configurations
(various mean positions of the gaussians in the first case)
to deduce the configuration of minimum energy.
In the most general case, the considerations of this section permit one to understand that, from the point of view of the KS potential, traditional DFT
fundamentally requires that the electron-nuclear correlation energy $E^{(ne)}_{XC}[\rho^{(n)}_{int},\rho^{(e)}_{int}]$ is neglected,
but not necessarily the clamped nuclei approximation.

Concerning the nuclear potential $v^{(n)}_S$, note that it is similar to the "one kind of particule" \textit{internal} KS potential 
(as c.m. correlations are included in $E^{(n)}_{XC}$),
with (using the notations of \cite{Mes09})
$$v^{int}(\vec{r})=v^{(n)}_{int}(\vec{r})+\int d\vec{r'} \rho^{(e)}_{int}(\vec{r'}) u^{(en)}(\vec{r}-\vec{r'}).$$

\subsubsection{Classical (point-like) nuclei.}
\label{par:tradDFT3}

A quantum treatment of the nuclei may be necessary for an accurate description of small molecules \cite{Gro95}
and certain molecular systems of intermediate size.
For larger molecular systems, a classical treatment of the nuclei is generally sufficient because the nuclei become
more and more localized in $\{\mathbf{r}\}$ and $\{\mathbf{p}\}$ spaces.
A further approximation is to suppose that the nuclei are perfectly localized in both spaces.
The appendix \ref{app:jacobi5} shows how to make this approximation properly in the case of self-bound systems.
We note $\{\mathbf{r}^{(ncl)}_i,\mathbf{p}^{(ncl)}_i\}$ the positions and momenta of the classical nuclei
in the c.m. frame, i.e. satisfying $\sum_{i=1}^{N^{(n)}} \mathbf{r}^{(ncl)}_i = 0$ and $\sum_{i=1}^{N^{(n)}} \mathbf{p}^{(ncl)}_i = 0$.
We pose $|\varphi^{(n)i}_{int}(\mathbf{r})|^2\rightarrow \delta (\mathbf{r}-\mathbf{r}^{(ncl)}_i)$ 
and $|\varphi^{(n)i}_{int}(\mathbf{p})|^2\rightarrow \delta (\mathbf{p}-\mathbf{p}^{(ncl)}_i)$,
so that $\rho^{(n)}_{int}(\mathbf{r})\rightarrow \sum_{i=1}^{N^{(n)}}\delta (\mathbf{r}-\mathbf{r}^{(ncl)}_i)$ 
and $\rho^{(n)}_{int}(\mathbf{p})\rightarrow \sum_{i=1}^{N^{(n)}}\delta (\mathbf{p}-\mathbf{p}^{(ncl)}_i)$.
The total internal energy (\ref{eq:action7}) becomes
(remember that only a Hartree self-interaction correction remains in $E^{(n)}_{XC}$ at the classical pointlike limit, see \S \ref{par:tradDFT1},
and that in the stationary case each $\mathbf{p}^{(ncl)}_i$ should be null)
\begin{eqnarray}
\lefteqn{E_{int}[\rho^{(e)}_{int},\{\mathbf{r}^{(ncl)}_i\}]= }
\nonumber\\
&& \sum_{i=1}^{N^{(e)}} (\varphi^{(e)i}_{int}|\frac{\vec{p}^2}{2m^{(e)}}|\varphi^{(e)i}_{int})
+ E^{(e)}_{H}[\rho^{(e)}_{int}] + E^{(e)}_{XC}[\rho^{(e)}_{int}] 
\nonumber\\
&& + \int d\vec{r} v^{(e)}_{int}(\vec{r}) \rho^{(e)}_{int}(\vec{r})
+ \sum_{i=1}^{N^{(n)}} \int d\vec{r} \rho^{(e)}_{int}(\vec{r}) u^{(en)}(\mathbf{r}-\mathbf{r}^{(ncl)}_i)
\nonumber\\
&& + \sum_{i>j=1}^{N^{(n)}} u^{(n)}(\mathbf{r}^{(ncl)}_i-\mathbf{r}^{(ncl)}_j)
+ \sum_{i=1}^{N^{(n)}} v^{(n)}_{int}(\mathbf{r}^{(ncl)}_i)
,
\label{eq:action6}
\end{eqnarray}
and the electronic KS potential (\ref{eq:varphi_i3}) becomes
\begin{eqnarray}
v^{(e)}_S(\mathbf{r}) &=&
\int d\vec{r'} \rho^{(e)}_{int}(\vec{r'}) u^{(e)}(\vec{r}-\vec{r'}) + U^{(e_{/e})}_{XC}[\rho^{(e)}_{int}](\vec{r})
\nonumber\\
&&+ v^{(e)}_{int}(\vec{r}) + \sum_{i=1}^{N^{(n)}} u^{(en)}(\mathbf{r}-\mathbf{r}^{(ncl)}_i)
,
\label{eq:varphi_i4}
\end{eqnarray}
which now depends parametrically on the nuclear positions $\{\mathbf{r}^{(ncl)}_i\}$.
Thus, the $\{\varphi^{(e)i}_{int}\}$ and $\rho^{(e)}_{int}$ also depend parametrically on the $\{\mathbf{r}^{(ncl)}_i\}$.
We recover traditional "one kind of particle" DFT equation \cite{Hoh64,Koh65,Dre90,Koh99,DFTLN}
with a classical nuclear background:
$$
v^{ext}(\vec{r})=v^{(e)}_{int}(\vec{r})+ \sum_{i=1}^{N^{(n)}} u^{(en)}(\mathbf{r}-\mathbf{r}^{(ncl)}_i).
$$

The nuclear KS equations
become obsolete in the stationary case.
Indeed, classical pointlike nuclei have no zero point motions,
so that they fall to the bottom of the potential well,
and the ground state of the whole molecule can be found by minimization of the total internal energy (\ref{eq:action6})
for various nuclear configurations $\{\mathbf{r}^{(ncl)}_i\}$.


\subsubsection{A method to improve the results of traditional DFT.}

The previous considerations permit us to set up a method to improve the results of traditional DFT in the molecular case:
\begin{enumerate}[(i)]
\item Start from the classical nuclei approximation described in \S \ref{par:tradDFT3},
with electrons described by "one kind of particle" DFT and parametrized by the nuclear positions.
This is what is done in most of practical calculations.
This is essentially suited for molecular systems whose nuclei are very localized, i.e. quite large molecular systems.
\item If this is not enough, for instance in the case of relatively small molecules where quantum effects associated with the nuclei play a role,
skip to the approximation described in \S \ref{par:tradDFT2}.
The nuclei are then treated quantum mechanically but the electron-nuclear correlation energy is neglected.
There remain the ($n$)-($n$) Hartree interaction with at least a self-interaction correction, 
until other more satisfactory nuclear exchange-correlation functionals (including the c.m. correlations) are available.
The electrons still satisfy "one kind of particle" traditional KS equations, but 
the corresponding orbitals are no longer parametrized by the nuclear positions.
\item If this is still not enough, for instance in cases where the electron-nuclei correlations play an important role,
add the electron-nuclear correlation energy
as described in \S \ref{par:tradDFT1}, using for instance the approximation proposed in \cite{Cha08}
(but further developements appear desirable).
The obtained equations go beyond "one kind of particle" traditional and internal KS equations,
because of the coupling correlation term $E^{(ne)}_{C}[\rho^{(n)}_{int},\rho^{(e)}_{int}]$.
\end{enumerate}

\section{On the application to atomic nuclei and mixtures of $^3$He and $^4$He droplets.}

When the masses of each species constituting the self-bound system are not very different
(for instance in the case of protons and neutrons in an atomic nucleus or mixtures of $^3$He and $^4$He droplets), 
one cannot do the approximations presented in the previous section, because the c.m. correlations couple all the particles, and the
complete formalism presented in \S \ref{par:intKS} has to be used.

One interest of the application of multicomponent internal DFT formalism to protons and neutrons
is to give a fundamental justification to the use of internal density functionals
for stationary mean-field like calculations of nuclei with effective interactions \cite{Ben03},
showing that there exists an ultimate \textit{local} potential which contains the c.m.\ correlations and permits one to 
reproduce the exact internal densities of the protons and neutrons
\footnote{
Protons and neutrons are both Fermions, so that the two functionals $E^{(l)}_{XC}$ should contain \textit{exchange} and correlations.
}.
It gives a more fundamental justification than the Hartree-Fock (HF) framework
to the stationary nuclear mean-field like calculations. Indeed, HF does not contain quantum correlations,
nor does it treat correctly the redundant coordinate problem, which introduces a spurious coupling between
internal properties and c.m. motion \cite{RS80,Schm01a}.
A way to overcome this problem in the stationary case is to perform projected HF
(projection before variation on c.m.\ momentum), which permits to restore 
Galilean invariance, but at the price of abandoning the independent-particle 
description \cite{Schm01a,Schm04,Pei62,Ben03} and at a large numerical cost.
The internal DFT formalism demonstrates that the c.m. correlations can be taken into account through a local potential,
\textit{a priori} numerically much less costly than the projection techniques.

We underline that, for nuclear systems, the $v^{(l)}_{int}$ are generally zero.
It is the dependence on the initial state that allows, in the time-dependent case, to describe for instance
the collision of two nuclei in the frame attached to the total c.m.\ of the nuclei \cite{Mes09-2}.

We mention that the point of view often adopted in nuclear physics
is that the protons and neutrons are the same kinds of particles, leading to isospin considerations \cite{RS80}.
From this point of view, it is rather the inclusion of the isospin in "one kind of particle" internal DFT 
that would give KS equations comparable to nuclear mean-field like calculations.

The application of multicomponent internal DFT to mixtures of $^3$He and $^4$He droplets
\footnote{
$^3$He are Fermions, so that the corresponding $E^{(l)}_{XC}$ functional should contain \textit{exchange} and correlations.
$^4$He are Bosons, so that the corresponding $E^{(l)}_{XC}$ functional should contain only correlations and the corresponding
orbitals $\varphi^{(l)i}_{int}$ should be identical (for condensates).
}
also permits us to give a fundamental justification to the mean-field like calculations done to describe those systems \cite{Bar06}.
Since the c.m. correlations are, to our knowledge, still not treated in existing calculations of Helium droplets,
this work opens the way to their inclusion.

%
%
\section{Conclusion.}

We have generalized the internal DFT formalism and Kohn-Sham scheme
to multicomponent self-bound systems, treating correctly their fundamental translational symmetry.
The formalism we obtained applies to the description of 
molecular systems where the nuclei are treated explicitly, atomic nuclei and mixtures of $^3$He and $^4$He droplets.
The main difference with traditional DFT is the explicit inclusion 
of the quantum center-of-mass correlations (due to the zero-point motion) in the functional.

The application to molecular systems where the nuclei are treated explicitly
permits us to clarify the approximations that underly traditional "one kind of particle" DFT and KS scheme
(i.e. electron-nuclei correlation energy neglected)
and why this formalism is essentially suited for the description of the electrons (only) in the frame attached to the c.m. of the nuclei.
We also set up a method to improve the results of traditional DFT in the case of molecular systems.

The application to atomic nuclei and mixtures of $^3$He and $^4$He droplets
provides a step towards a fundamental justification to the use of effective interactions that are functionals of the
one-body densities of each species in mean-field like calculations.

Finally, we mention the questions that remain open:
\begin{enumerate}[(i)]
\item The search for a general functional that describes the c.m.\ correlations and leads to a local KS potential is continuing.
It would give a numerically advantageous 
alternative to the projection techniques used in nuclear physics to restore Galilean invariance.
It could also be used to improve the description of Helium droplets and nuclei in molecular systems.
\item In the case of molecular systems, it appears desirable to find more satisfactory functional forms for the nucleus-nucleus correlation energy $E^{(n)}_{XC}[\rho^{(n)}_{int}]$
than the simple self-interaction correction to the Hartree energy obtained through the crude classical (pointlike) approximation.
One way could be to use semi-classical approximations.
It also appears desirable to pursue the search of functionals for the electron-nuclear correlation energy $E^{(ne)}_{C}[\rho^{(n)}_{int},\rho^{(e)}_{int}]$.
\item The generalization this work to the time-dependent case is not completely trivial
and is under investigation, following the considerations of Refs.~\cite{Mes09-2} and \cite{Li86}.
\item
\JMcomm{
The question of the non-interacting v-representability according to the number of orbitals introduced in the KS scheme
merits further study.
}
\item It would be interesting to apply the same reasoning
to rotational invariance to formulate the theory in term of the internal density with respect
to the body-fixed frame (sometimes also called "intrinsic" one-body density \cite{Gir08a}).
Indeed, even if this density is not directly observable experimentally,
there is some indirect experimental evidence for the existence of rotational symmetry breaking states
in nuclear rotational bands.
Such a work would certainly shed some light on the symmetry breaking question.
\end{enumerate}


\subsection*{Acknowledgments.}

The author is particularly grateful to M. Bender, E. Suraud and T. Duguet for numerous enlightening discussions and
careful reading of the manuscript, and to John Donohue for careful reading of the manuscript.
The author gratefully acknowledges the Conseil R\'egional d'Aquitaine for financial support and
the Centre d'Etudes Nucl\'eaires de Bordeaux-Gradignan for warm hospitality.


\begin{appendix}

\section{Jacobi coordinates for particles of different kinds.}
\label{app:jacobi2}

We consider a system of $N$ particles of different kinds.
We start from the set of conjugate laboratory coordinates $\{\mathbf{r}_i\}$ and $\{\mathbf{p}_i\}$, associated to the masses $\{m_i\}$, with $i=1,...,N$. The Jacobi coordinates are ($\alpha=1,...,N-1$):
\begin{enumerate}[(i)]
\item the c.m. position $\mathbf{R}=\frac{1}{\sum_{i=1}^N m_i}\sum_{i=1}^N m_i \mathbf{r}_i$,
conjugated to the total momentum $\mathbf{P}=\sum_{i=1}^N \mathbf{p}_i$,
\item the relative positions (relative to the c.m. of the $i-1$ other particles) 
$\mathbf{\xi}_\alpha = \mathbf{r}_{\alpha+1} - \frac{1}{\sum_{i=1}^\alpha m_i}\sum_{i=1}^\alpha m_i \mathbf{r}_i$,
conjugate to the relative momenta\\
$\mathbf{\tau}_\alpha = \frac{1}{\sum_{i=1}^{\alpha+1}m_i } \Big( \mathbf{p}_{\alpha+1}\sum_{i=1}^\alpha m_i - m_{\alpha+1} \sum_{i=1}^\alpha \mathbf{p}_i \Big)$.
\end{enumerate}
If we note $\sum_{i=1}^N m_i=M$ the total mass and \\
$\mu_\alpha= \frac{m_{\alpha+1}\sum_{i=1}^\alpha m_i}{\sum_{i=1}^{\alpha+1}m_i}$ the relative mass, we can decompose the kinetic energy as 
$\sum_{i=1}^N \frac{\mathbf{p}_i^2}{2m_i}=\frac{\mathbf{P}^2}{2M} + \sum_{\alpha=1}^{N-1}\frac{\mathbf{\tau}_\alpha^2}{2\mu_\alpha}$.

\section{Simplification of the Jacobi coordinates when one kind of particle is much heavier than the other.}
\label{app:jacobi3}

We suppose that $m^{(1)}>>m^{(2)}$.
Then, if $N^{(2)}$ is not much larger than $N^{(1)}$ and/or $\vec{R}^{(2)}$ is not very far away from $\vec{R}^{(1)}$,
which is the case for molecular systems, one can simplify (\ref{eq:cm}) in
\begin{eqnarray}
\vec{R}&=&\frac{1}{N^{(1)}} \sum_{i=1}^{N^{(1)}}\vec{r}^{(1)}_j = \mathbf{R}^{(1)},
\end{eqnarray}
so that the c.m. of the whole system coincides with the c.m. $\mathbf{R}^{(1)}$ of the particles $(1)$ only.
Moreover, Jacobi coordinates (\ref{eq:jacobi}) and reduced masses (\ref{eq:red_mass}) can be simplified in
\begin{eqnarray}
&&for \hspace{3mm}\alpha\in[1;N^{(1)}-1]: 
\nonumber\\
&& \hspace{1.5mm}\mathbf{\xi}_{\alpha} = 
\vec{r}^{(1)}_{\alpha+1}-\frac{1}{\alpha}\sum_{i=1}^{\alpha}\vec{r}^{(1)}_i = \mathbf{\xi}_{\alpha}^{(1)}
\hspace{1.5mm} and\hspace{1.5mm} \mu_\alpha = \frac{\alpha}{\alpha+1}m^{(1)},
\nonumber\\
&&for \hspace{3mm}i\in[1;N^{(2)}]:
\nonumber\\
&& \hspace{1.5mm}\mathbf{\xi}_{N^{(1)}-1+i} = 
\vec{r}^{(2)}_{i}-\vec{R}^{(1)} = \mathbf{r'}_i^{(2)}
\hspace{1.5mm} and\hspace{1.5mm} \mu_{N^{(1)}-1+i} = m^{(2)}
\nonumber,
\end{eqnarray}
where the ($N^{(1)}-1$) coordinates $\mathbf{\xi}_{\alpha}^{(1)}$ appear as Jacobi coordinates specific to the particles of kind $(1)$ only,
and the remaining $N^{(2)}$ coordinates $\mathbf{r'}^{(2)}_i$ appear naturally as the coordinates of each $N^{(2)}$ single particles $(2)$ in the c.m. frame.
Thus, we see that we obtain an equivalent result if we apply Jacobi coordinates to the heavy particles only
and if we describe each light particle by its coordinates 
in the frame attached to the c.m. of the heavy particles.
As a consequence, the redundant coordinate problem, and thus the c.m.\ correlations, concerns only the heavy particles.
This explains why those correlations should not appear in the exchange-correlation energy functional of the light particles
(for instance the electrons in a molecule).


\section{Exchange-correlation functionals when one kind of particle is much heavier than the other.}
\label{app:jacobi4}

In the molecular case, as $\vec{R}=\mathbf{R}^{(n)}$, for any \JMcomm{function} $f(\vec{r}^{(e)}_{1}, \ldots , \vec{r}^{(e)}_{N^{(e)}})$
that depends on the electronic coordinates only, we have
\begin{eqnarray}
\lefteqn{I_f}
\nonumber\\
&=& \int \! d\vec{r}^{(n)}_1 \cdots d\vec{r}^{(n)}_{N^{(n)}} d\vec{r}^{(e)}_1 \cdots d\vec{r}^{(e)}_{N^{(e)}}
\delta(\mathbf{R})
\nonumber\\
&\times& |\psi_{int}(\vec{r}^{(n)}_1, \ldots , \vec{r}^{(n)}_{N^{(n)}};\vec{r}^{(e)}_1, \ldots , \vec{r}^{(e)}_{N^{(e)}})|^2 
f(\vec{r}^{(e)}_{1}, \ldots , \vec{r}^{(e)}_{N^{(e)}})
\nonumber\\
&=&
\int d\vec{r}^{(e)}_1 \cdots d\vec{r}^{(e)}_{N^{(e)}}
f(\vec{r}^{(e)}_{1}, \ldots , \vec{r}^{(e)}_{N^{(e)}})
|\psi_{int}^{(e)}(\vec{r}^{(e)}_{1}, \ldots , \vec{r}^{(e)}_{N^{(e)}})|^2
\nonumber
,
\end{eqnarray}
%
where we have defined the local part of the $N^{(e)}$-body electronic density as
\begin{eqnarray}
&&|\psi_{int}^{(e)}(\vec{r}^{(e)}_{1}, \ldots , \vec{r}^{(e)}_{N^{(e)}})|^2= 
\int \! d\vec{r}^{(n)}_1 \cdots d\vec{r}^{(n)}_{N^{(n)}} \delta(\mathbf{R}^{(n)})
\nonumber\\
&&\hspace{12mm}\times
|\psi_{int}(\vec{r}^{(n)}_1, \ldots , \vec{r}^{(n)}_{N^{(n)}};\vec{r}^{(e)}_1, \ldots , \vec{r}^{(e)}_{N^{(e)}})|^2
\label{eq:psi_e}
.
\end{eqnarray}
This makes it clear that $I_f$ is no longer explicitly affected by (or no longer contains explicitly) the c.m. correlations.
Indeed, the $\delta(\mathbf{R}^{(n)})$ does not directly affect the integral over the electronic coordinates.
The c.m. correlations are included only implicitly through $|\psi^{(e)}_{int}|^2$.

More generally, we consider an operator $\hat{F}$ that acts on the electronic coordinates only
(we write $\hat{F}(\vec{r}^{(e)}_{1}, \ldots , \vec{r}^{(e)}_{N^{(e)}})$ in $\{\mathbf{r}\}$ representation). We find
\begin{widetext}
\begin{eqnarray}
I_{\hat{F}} &=& \int \! d\vec{r}^{(n)}_1 \cdots d\vec{r}^{(n)}_{N^{(n)}} d\vec{r}^{(e)}_1 \cdots d\vec{r}^{(e)}_{N^{(e)}}
\delta(\mathbf{R})\times
\nonumber\\
&&\quad\quad\quad\quad\quad\quad
\psi_{int}^*(\vec{r}^{(n)}_1, \ldots , \vec{r}^{(n)}_{N^{(n)}};\vec{r}^{(e)}_1, \ldots , \vec{r}^{(e)}_{N^{(e)}}) 
\hat{F}(\vec{r}^{(e)}_{1}, \ldots , \vec{r}^{(e)}_{N^{(e)}})
\psi_{int}(\vec{r}^{(n)}_1, \ldots , \vec{r}^{(n)}_{N^{(n)}};\vec{r}^{(e)}_1, \ldots , \vec{r}^{(e)}_{N^{(e)}})
\nonumber\\
&=&
\int d\vec{r}^{(e)}_1 \cdots d\vec{r}^{(e)}_{N^{(e)}}
\psi_{int}^{(e)*}(\vec{r}^{(e)}_{1}, \ldots , \vec{r}^{(e)}_{N^{(e)}})
\hat{F}(\vec{r}^{(e)}_{1}, \ldots , \vec{r}^{(e)}_{N^{(e)}})
\psi_{int}^{(e)}(\vec{r}^{(e)}_{1}, \ldots , \vec{r}^{(e)}_{N^{(e)}})
,
\end{eqnarray}
\end{widetext}
where $\psi_{int}^{(e)}$ is defined such as
\begin{widetext}
\begin{eqnarray}
&&\psi_{int}^{(e)*}(\vec{r}^{(e)}_{1}, \ldots , \vec{r}^{(e)}_{N^{(e)}})
\hat{F}(\vec{r}^{(e)}_{1}, \ldots , \vec{r}^{(e)}_{N^{(e)}})
\psi_{int}^{(e)}(\vec{r}^{(e)}_{1}, \ldots , \vec{r}^{(e)}_{N^{(e)}})=
\nonumber\\
&&\int \! d\vec{r}^{(n)}_1 \cdots d\vec{r}^{(n)}_{N^{(n)}} \delta(\mathbf{R}^{(n)})
\psi_{int}^*(\vec{r}^{(n)}_1, \ldots , \vec{r}^{(n)}_{N^{(n)}};\vec{r}^{(e)}_1, \ldots , \vec{r}^{(e)}_{N^{(e)}}) 
\hat{F}(\vec{r}^{(e)}_{1}, \ldots , \vec{r}^{(e)}_{N^{(e)}})
\psi_{int}(\vec{r}^{(n)}_1, \ldots , \vec{r}^{(n)}_{N^{(n)}};\vec{r}^{(e)}_1, \ldots , \vec{r}^{(e)}_{N^{(e)}})
.
\nonumber
\end{eqnarray}
\end{widetext}
This definition complies with (\ref{eq:psi_e}) (when $\hat{F}\rightarrow f$).
As above, we see that $I_{\hat{F}}$ is not explicitly affected by the c.m. correlations
(those correlations are included only implicitly through the "electronic wave function" $\psi_{int}^{(e)}$).

All this makes it clear, from a complementary point of view of that presented in appendix \ref{app:jacobi3},
that the c.m. correlations should no longer appear explicitly in the electronic exchange-correlation functional (\ref{eq:Axc}).
Indeed it can be rewritten, with the previous notation
\begin{widetext}
\begin{eqnarray}
\label{eq:Exc_e}
E^{(e)}_{XC}
&=& \frac{1}{2} \int \! d\vec{r} \, d\vec{r'} \, 
      \Big[ \gamma^{(e)}_{int}(\vec{r},\vec{r'}) - \rho^{(e)}_{int}(\vec{r}) \, \rho^{(e)}_{int}(\vec{r'}) \Big] \, 
      u^{(e)}(\vec{r}-\vec{r'})
\\
&&    + \sum_{i=1}^{N^{(e)}} \Big[
      \int \! d\vec{r}^{(e)}_1 \cdots d\vec{r}^{(e)}_{N^{(e)}} \,
      \psi_{int}^{(e)*}(\vec{r}^{(e)}_1, \cdots, \vec{r}^{(e)}_{N^{(e)}})
      \frac{\mathbf{p}^{(e)2}_i}{2m^{(e)}}
      \psi_{int}^{(e)}(\vec{r}^{(e)}_1, \cdots, \vec{r}^{(e)}_{N^{(e)}})
      - (\varphi^{(e)i}_{int}|\frac{\vec{p}^{2}}{2m^{(e)}}|\varphi^{(e)i}_{int}) \Big]
\nonumber
,
\end{eqnarray}
\end{widetext}
whose form is similar to that of traditional DFT \cite{Dre90}.
Doing the same kind of reasoning for the nuclei, we obtain
\begin{widetext}
\begin{eqnarray}
\label{eq:Exc_n}
E^{(n)}_{XC}
&=& \frac{1}{2} \int \! d\vec{r} \, d\vec{r'} \, 
      \Big[ \gamma^{(n)}_{int}(\vec{r},\vec{r'}) - \rho^{(n)}_{int}(\vec{r}) \, \rho^{(n)}_{int}(\vec{r'}) \Big] \, 
      u^{(n)}(\vec{r}-\vec{r'})
\\
&&    + \sum_{i=1}^{N^{(n)}} \Big[
      \int \! d\vec{r}^{(n)}_1 \cdots d\vec{r}^{(n)}_{N^{(n)}} \,
      \delta(\mathbf{R}^{(n)})
      \psi_{int}^{(n)*}(\vec{r}^{(n)}_1, \cdots, \vec{r}^{(n)}_{N^{(n)}})
      \frac{\mathbf{p}^{(n)2}_i}{2m^{(n)}}
      \psi_{int}^{(n)}(\vec{r}^{(n)}_1, \cdots, \vec{r}^{(n)}_{N^{(n)}})
      - (\varphi^{(n)i}_{int}|\frac{\vec{p}^{2}}{2m^{(n)}}|\varphi^{(n)i}_{int}) \Big]
\nonumber
,
\end{eqnarray}
\end{widetext}
which explicitly contains all the c.m. correlations and
whose form is similar to that of internal DFT \cite{Mes09}.

\section{The classical (point-like) limit in the internal DFT formalism.}
\label{app:jacobi5}

To lighten the notation and since it is sufficient for the considerations of this paper, we consider only one kind of particle in this section.
The generalization to many kinds of particles is straightforward.

\subsection{Classical limit for the densities.}

The classical (point-like) limit for a N-body density $|\tilde\psi|^2$ is usually obtained by assuming
\begin{eqnarray}
|\tilde\psi(\vec{r}_{1}, \ldots , \vec{r}_{N})|^2
\rightarrow \frac{1}{N!} \sum_{P} \Pi_{i=1}^{N} \rho^{(cl)i} (\mathbf{r}_{P(i)})
,
\end{eqnarray}
where the $\{\vec{r}_{i}\}$ are the coordinates of the $N$ particles,
$P$ the possible permutations of those coordinates
and
\begin{eqnarray}
\rho^{(cl)i} (\mathbf{r})=
\delta ( \mathbf{r}-\mathbf{r}^{(cl)}_i )
\end{eqnarray}
are the classical densities of each particle $i$ (the $\mathbf{r}^{(cl)}_i$ being their classical positions).
Of course, this approximation breaks the translational invariance, so that it can be fundamentally justified
only for densities $|\tilde\psi|^2$ that are \textit{not} translationally invariant (and are symmetric under the exchange of two particles, of course).
It thus \textit{cannot} be fundamentally justified for the laboratory N-body density $|\psi|^2$ of a self-bound system,
nor for its internal N-body density $|\psi_{int}|^2$, expressed with the particles coordinates $\{\vec{r}_{i}\}$,
because they are both translationally invariant.

For self-bound systems, it is in fact for the c.m. frame N-body "density"
$\delta(\mathbf{R}) |\psi_{int}(\vec{r}_{1}, \ldots , \vec{r}_{N})|^2$
that the classical approximation can fundamentally be justified.
Indeed, this "density" is obviously not translationally invariant
(the $\delta(\mathbf{R})$ fixes the c.m. in position space and amounts to move in the c.m. frame)
and is symmetric under the exchange of two particles
(as $|\psi_{int}|^2$ satisfies this symmetry).
It is non null only for the $\{\vec{r}_{i}\}$ that satisfy $\mathbf{R}=\sum_{i=1}^N \vec{r}_{i} = 0$
(i.e. the $\{\vec{r}_{i}\}$ become the c.m. frame coordinates).

The same way, for a classical system described in the c.m. frame,
$\frac{1}{N!} \sum_{P} \Pi_{i=1}^{N} \rho^{(cl)}_i (\mathbf{r}_{P(i)})$
is non null only for the $\{\vec{r}^{(cl)}_{i}\}$ that satisfy $\sum_{i=1}^N \vec{r}^{(cl)}_{i} = 0$.
Thus, for self-bound systems,
the proper way to do the classical approximation is to replace
\begin{eqnarray}
\label{eq:class}
\delta(\mathbf{R}) |\psi_{int}(\vec{r}_{1}, \ldots , \vec{r}_{N})|^2
\rightarrow \frac{1}{N!} \sum_{P} \Pi_{i=1}^{N} \rho^{(cl)i}_{int} (\mathbf{r}_{P(i)})
,
\end{eqnarray}
with
\begin{eqnarray}
\sum_{i=1}^N \vec{r}^{(cl)}_{i} = 0
\quad\quad and \quad\quad 
\rho^{(cl)i}_{int} (\mathbf{r})=
\delta ( \mathbf{r}-\mathbf{r}^{(cl)}_i )
.
\end{eqnarray}
Inserting this approximation in the definitions of \S \ref{par:def}, we obtain the classical (point-like) limits of the one and two-body internal densities
\begin{eqnarray}
&& \rho^{(cl)}_{int}(\mathbf{r})= \sum_{i=1}^N \rho^{(cl)i}_{int} (\mathbf{r})
,
\\
&& \gamma^{(cl)}_{int}(\mathbf{r},\mathbf{r}')= \rho^{(cl)}_{int}(\mathbf{r})\rho^{(cl)}_{int}(\mathbf{r}')
- \sum_{i=1}^N \rho^{(cl)i}_{int} (\mathbf{r}) \rho^{(cl)i}_{int} (\mathbf{r}')
\nonumber
.
\end{eqnarray}
%

\subsection{Classical limit for the kinetic energy terms.}

We define the one-body density in $\{\mathbf{p}\}$ space of the interacting system as
$$
\rho_{int}(\mathbf{p}) = N \int d\mathbf{p}_1 \dots d\mathbf{p}_N \delta(\mathbf{P})
|\psi_{int}(\vec{p}_{1}, \ldots , \vec{p}_{N})|^2 \delta(\vec{p}_{i}-\vec{p})
,
$$
where $\mathbf{P}=\sum_{i=1}^N \vec{p}_{i}$ is fixed to zero, so that $\rho_{int}$ is defined in the c.m. frame.
The interacting internal kinetic energy can then be written
$$
(\psi_{int}|\sum_{\alpha=1}^{N-1} \frac{\tau_\alpha^2}{2\mu_\alpha}|\psi_{int})
=
\int d\mathbf{p} \hspace{1mm}\frac{\mathbf{p}^2}{2m}\hspace{1mm} \rho_{int}(\mathbf{p})
.
$$
The one-body density in $\{\mathbf{p}\}$ space of the non-interacting (KS) system is defined as \cite{Dre90}
$$
\rho^{S}_{int}(\mathbf{p}) = \int d\mathbf{r} d\mathbf{r}'
\exp^{i\mathbf{p}(\mathbf{r}-\mathbf{r}')/\hbar}\sum_{i=1}^N \varphi_{int}^i(\mathbf{r})\varphi_{int}^{i*}(\mathbf{r'})
,
$$
so that the kinetic energy of the non-interacting system can be written
$$
\sum_{i=1}^N(\varphi^{i}_{int}|\frac{\vec{p}^{2}}{2m}|\varphi^{i}_{int})
=
\int d\mathbf{p}\hspace{1mm} \frac{\mathbf{p}^2}{2m}\hspace{1mm} \rho^S_{int}(\mathbf{p})
.
$$
Even if the the one-body densities in the $\{\mathbf{r}\}$ space of the interacting and non-interacting systems are the same,
it is not the case for the  one-body densities in the $\{\mathbf{p}\}$ space, i.e. $\rho_{int}(\mathbf{p})\ne\rho^{S}_{int}(\mathbf{p})$
(the difference is due to the quantum correlations).
Thus, in the general case, the difference between interacting and non-interacting internal kinetic energies that appears in $E_{XC}$, 
see second line of Eq.~(\ref{eq:Axc}), is not null:
\begin{eqnarray}
\lefteqn{(\psi_{int}|\sum_{\alpha=1}^{N-1} \frac{\tau_\alpha^2}{2\mu_\alpha}|\psi_{int})
-\sum_{i=1}^N(\varphi^{i}_{int}|\frac{\vec{p}^{2}}{2m}|\varphi^{i}_{int}) =}
\nonumber\\
&&\hspace{10mm}\int d\mathbf{p}\hspace{1mm} \frac{\mathbf{p}^2}{2m}\hspace{1mm} \Big( \rho_{int}(\mathbf{p}) - \rho^S_{int}(\mathbf{p}) \Big)
\ne 0
.
\end{eqnarray}
But, since the mean total momenta of the interacting and non-interacting systems should be equal, the classical (point-like) limit
(obtained supposing that the system is
perfectly localized in both $\{\mathbf{r}\}$ and $\{\mathbf{p}\}$ spaces) gives
$$
\rho_{int}(\mathbf{p})
= \rho^S_{int}(\mathbf{p})
= \sum_{i=1}^N \delta ( \mathbf{p}-\mathbf{p}^{(cl)}_i )
,
$$
where the $\mathbf{p}^{(cl)}_i$ are the classical momenta of the particles in the c.m. frame, which satisfy $\sum_{i=1}^N \vec{p}^{cl}_{i}=0$
(in the stationary case, $\vec{p}^{cl}_{i}=0$ of course).
Thus, in the classical limit, the "kinetic energy exchage-correlations" part of $E_{XC}$ (i.e. the second line of Eq.~(\ref{eq:Axc})),
which contains explicitly the c.m. correlations in $\{\mathbf{r}\}$ representation, disappears.

\end{appendix}

\end{document}